\documentclass{ws-ijmpa}
\usepackage[super,compress]{cite}
\def\Journal#1#2#3#4{{#1} {\bf #2}, #3 (#4)}

\def\NIMA{\em Nucl. Instrum. Methods A}

\def\PLB{\em Phys. Lett.  B}
\def\PRL{\em Phys. Rev. Lett.}
\def\PRD{\em Phys. Rev. D}

\def\GaC{\em Gravitation and Cosmology}
\def\GaCS{\em Gravitation and Cosmology Suppl.}

\def\PAN{\em Phys.Atom.Nucl.}

\def\APJ{\em Astrophys. J.}
\def\SCI{\em Science}
\def\MPLA{\em Mod. Phys. Lett. A}
\def\IJTP{\em Int. J. Theor. Phys.}
\def\IJMPA{\em Int. J. Mod. Phys. A}
\def\IJMPD{\em Int. J. Mod. Phys.  D}
\def\NJP{\em New J. of Phys.}
\def\ARAA{\em Ann. Rev. Astron. Astrophys.}
\def\AIPCP{\em AIP Conf. Proc.}
\def\JHEP{\em JHEP}
\def\JCAP{\em JCAP}

\def\JPCS{\em J. Phys. Conf. Ser.}
\def\BWP{\em Bled Workshops in Physics}
\def\s{{\,\rm s}}
\def\g{{\,\rm g}}
\def\eV{\,{\rm eV}}
\def\keV{\,{\rm keV}}
\def\MeV{\,{\rm MeV}}
\def\GeV{\,{\rm GeV}}
\def\TeV{\,{\rm TeV}}
\def\sv{\left<\sigma v\right>}
\def\({\left(}
\def\){\right)}
\def\cm{{\,\rm cm}}
\def\K{{\,\rm K}}
\def\kpc{{\,\rm kpc}}
\def\beq{\begin{equation}}
\def\eeq{\end{equation}}
\def\bea{\begin{eqnarray}}
\def\eea{\end{eqnarray}}
\begin{document}

\markboth{M.Yu.Khlopov}
{Fundamental Particle Structure in the Cosmological Dark Matter}

%
\catchline{}{}{}{}{}
%

\title{FUNDAMENTAL PARTICLE STRUCTURE IN THE COSMOLOGICAL DARK MATTER
}

\author{MAXIM KHLOPOV}

\address{Centre for Cosmoparticle Physics "Cosmion"  \\
National Research Nuclear University "Moscow Engineering Physics Institute", 115409 Moscow, Russia \\
APC laboratory 10, rue Alice Domon et L\'eonie Duquet \\75205
Paris Cedex 13, France\\
khlopov@apc.univ-paris7.fr}

\maketitle

\begin{history}
\received{Day Month Year}
\revised{Day Month Year}
\end{history}

\begin{abstract}
The nonbaryonic dark matter of the Universe is assumed to consist of
new stable forms of matter. Their stability reflects symmetry of micro world and mechanisms of its symmetry breaking.
Particle candidates for cosmological
dark matter are lightest particles that bear new conserved quantum numbers. Dark matter particles may represent ideal gas of non-interacting particles. Self-interacting dark matter weakly or superweakly coupled to ordinary matter is also possible, reflecting nontrivial pattern of particle symmetry in the hidden sector of particle theory. In the early Universe the structure of particle symmetry breaking gives rise to cosmological phase transitions, from which macroscopic cosmological defects or primordial nonlinear structures can be originated. Primordial black holes (PBHs) can be not only a candidate for dark matter, but also represent a universal probe for super-high energy physics in the early Universe. Evaporating PBHs turn to be a source of even superweakly interacting particles, while clouds of massive PBHs can serve as a nonlinear seeds for galaxy formation. The observed broken symmetry of the three known families may provide a simultaneous solution for the problems of the mass of neutrino and strong CP violation in the unique framework of models of horizontal unification. Dark matter candidates can also appear in the new families of quarks and leptons and the existence of new stable charged leptons and quarks is possible, hidden in elusive "dark atoms". Such possibility, strongly restricted by
 the constraints on anomalous isotopes of light elements, is not excluded in scenarios that predict stable double charged particles.
The excessive -2 charged particles are bound in these scenarios with
primordial helium in O-helium "atoms", maintaining specific
nuclear-interacting form of the dark matter, which may
provide an interesting solution for the puzzles of the direct dark matter searches. In the context of cosmoparticle physics, studying fundamental relationship of micro- and macro- worlds, the problem of cosmological dark matter
implies cross disciplinary theoretical, experimental and observational studies for its solution.
\end{abstract}

\ccode{PACS numbers:12.60.Cn,98.90.+s,12.60.Nz,14.60.Hi,26.35.+c,36.90.+f,03.65.Ge}
\section{Introduction}
The convergence of the frontiers of our knowledge in micro- and
macro worlds leads to the wrong circle of problems, illustrated by
the mystical Ouroboros (self-eating-snake). The Ouroboros puzzle may
be formulated as follows: {\it The theory of the Universe is based
on the predictions of particle theory, that need cosmology for their
test}. Cosmoparticle physics
\cite{ADS,MKH,book,book3,newBook,Khlopov:2004jb,bled} offers the way out of
this wrong circle. It studies the fundamental basis and mutual
relationship between micro-and macro-worlds in the proper
combination of physical, astrophysical and cosmological signatures.
Some aspects of this relationship, which arise in the
problem of cosmological Dark Matter (DM), is the subject of this
review.

 Extensions of the standard model
imply new symmetries and new particle states. The respective
symmetry breaking induces new fundamental physical scales in
particle theory. In particle theory Noether's theorem relates the exact symmetry to
conservation of respective charge. If the symmetry is strict, the charge is strictly conserved. The lightest particle, bearing this charge, is
stable. It gives rise to the fundamental relationship between dark matter
candidates and particle symmetry beyond the Standard model.

If the symmetry is broken, the mechanism of the symmetry breaking implies restoration
of the symmetry at high temperatures and densities. Such high temperatures and densities
should have naturally arisen at the early stages of cosmological evolution.
It makes Big Bang Universe natural laboratory of particle physics, not only due to
possibility of creation of hypothetical particles in the early Universe, but also owing to
reflection of the hierarchy of particle symmetry breaking in cosmological phase transitions.

In the old Big Bang scenario  cosmological expansion and its initial
conditions were given {\it a priori} \cite{Weinberg,ZNSEU}. In the
modern cosmology expansion of Universe and its initial conditions
are related to inflation
\cite{Star80,Guthinfl,Linde:1981mu,Albrecht,Linde:1983gd},
baryosynthesis and nonbaryonic dark matter (see review in
Refs. \citen{Lindebook,Kolbbook,Rubakovbook1,Rubakovbook2}). The global properties of
the Universe as well as the origin of its large scale structure
are considered as the result of the process of inflation. The matter content of the modern
Universe is also originated from the physical processes: the
baryon density is the result of baryosynthesis and the nonbaryonic
dark matter represents the relic species of physics beyond the Standard model.
Here we would like to outline some
nontrivial forms of relationship between the cosmological problem of dark matter with
the fundamental symmetry of particle world.

According to the modern cosmology, the dark matter, corresponding to
$\sim 25\%$ of the total cosmological density, is nonbaryonic and
consists of new stable forms of matter. These forms of matter (see e.g. Refs. \citen{book,Cosmoarcheology,bled,Bled07,newBook,bertone}
for review and reference) should
be stable, saturate the measured dark matter density and decouple
from plasma and radiation at least before the beginning of matter
dominated stage. The easiest way to satisfy these conditions is to
involve neutral elementary weakly interacting particles. However it
is not the only particle physics solution for the dark matter
problem and more evolved models of the physical nature of dark matter are
possible.

Formation of the Large Scale Structure of the Universe from
small initial density fluctuations is one of the most
important reasons for the {\it nonbaryonic} nature of the dark matter
that is decoupled from matter and radiation and provides the effective growth of these fluctuations before recombination. It implies dark matter candidates from the physics beyond the Standard model (see Refs. \citen{bertone,LSSFW,Gelmini,Aprile:2009zzd,Feng:2010gw} for recent review). On the other hand, the initial density fluctuations, coming from the very early Universe are also originated from physics beyond the Standard model. In the present review we give some examples, linking the primordial seeds of galaxy formation to effects of particle symmetry breaking at very high energies.

Here we don't touch the exciting problems of the possible nature of dark matter related with extra dimensions and brane cosmology, but even in the case of our 1+3 dimensional space-time we find a lot of examples of nontrivial cosmological reflection of fundamental particle structure.

In the Section \ref{pattern} we present examples of cosmological pattern of fundamental particle symmetry: from various types of stable particle candidates for dark matter to primordial nonlinear structures, relics of phase transitions in the very early Universe. We then pay special attention to primordial black holes as a universal theoretical probe for new physics in the very early Universe (Section \ref{pbhs}). We give an example of a possibility to incorporate various types of dark matter within a unique framework of broken gauge symmetry of the three known families as well as discuss a possibility for stable charged species of new quarks and leptons to form dark matter, hidden in neutral dark atoms. In Section \ref{ohe} we consider specific form of O-helium (OHe) dark atoms that consist of heavy -2 charged heavy lepton-like particle surrounded by helium nuclear shell. The proof of qualitative advantages of this OHe scenario implies strict quantum mechanical solution of the problem of OHe interaction with nuclei. The conclusive Section \ref{Discussion} considers cosmological probes of fundamental particle structure in the context of cosmoparticle physics, studying fundamental relationship of micro- and macro- worlds.

\section{Cosmological pattern of particle physics}
 \label{pattern}
Let's specify in more details the set of links between fundamental
particle properties and their cosmological effects.

Most of the known particles are unstable. For a particle with the
mass $m$ the particle physics time scale is $t \sim 1/m$
\footnote{Here and further, if it isn't specified otherwise we use the units $\hbar=c=k=1$}, so in
particle world we refer to particles with lifetime $\tau \gg 1/m$
as to metastable. To be of cosmological significance in the Big Bang Universe metastable
particle should survive after the temperature of the Universe $T$
fell down below $T \sim m$, what means that the particle lifetime
should exceed $t \sim (m_{Pl}/m) \cdot (1/m)$. Such a long
lifetime should find reason in the existence of an (approximate)
symmetry. From this viewpoint, cosmology is sensitive to the most
fundamental properties of microworld, to the conservation laws
reflecting strict or nearly strict symmetries of particle theory.

So, electron is absolutely
stable owing to the conservation of electric charge, while the stability of proton is conditioned by the
conservation of baryon charge. The stability of ordinary matter is
thus protected by the conservation of electric and baryon charges,
and its properties reflect the fundamental physical scales of
electroweak and strong interactions. Indeed, the mass of electron
is related to the scale of the electroweak symmetry breaking,
whereas the mass of proton reflects the scale of QCD confinement.

The set of new
fundamental particles, corresponding to the new strict symmetry,
is then reflected in the existence of new stable particles, which
should be present in the Universe and taken into account in the
total energy density.

However, there
is no strict symmetry between various quarks and leptons. The
symmetry breaking implies the difference in particle masses. The
particle mass spectrum reflects the hierarchy of symmetry breaking.

The mechanism of spontaneous breaking of particle symmetry also has
cosmological impact. Heating of the condensed matter leads to
restoration of its symmetry. When the heated matter cools down,
phase transition to the phase of broken symmetry takes place. In the
course of the phase transitions, corresponding to given type of
symmetry breaking, topological defects can form. One can directly
observe formation of such defects in liquid crystals or in
superfluid He. In the same manner the mechanism of spontaneous
breaking of particle symmetry implies restoration of the underlying
symmetry in the early Universe at high temperatures.
When temperature decreases in the course of cosmological
expansion, transitions to the phase of broken symmetry  can lead,
depending on the symmetry breaking pattern, to formation of
topological defects in very early Universe. Defects can represent
new forms of stable particles (as it is in the case of magnetic
monopoles \cite{t'Hooft,polyakov,kz,Priroda,preskill,SAOmonop}), or
extended structures, such as cosmic strings \cite{zv1,zv2} or cosmic
walls \cite{okun}.

\subsection{Cosmoarcheology of new physics}
 \label{Cosmoarcheology}
Physics, underlying inflation,
baryosynthesis and dark matter, is referred to the extensions of
the standard model, and the variety of such extensions makes the
whole picture in general ambiguous. However, in the framework of
each particular physical realization of inflationary model with
baryosynthesis and dark matter the corresponding model dependent
cosmological scenario can be specified in all the details. In such
scenario the main stages of cosmological evolution, the structure
and the physical content of the Universe reflect the structure of
the underlying physical model. The latter should include with
necessity the standard model, describing the properties of
baryonic matter, and its extensions, responsible for inflation,
baryosynthesis and dark matter. In no case the cosmological impact
of such extensions is reduced to reproduction of these three
phenomena only. The nontrivial path of cosmological evolution,
specific for each particular implementation of inflational model with
baryosynthesis and nonbaryonic dark matter, always contains some
additional model dependent cosmologically viable predictions,
which can be confronted with astrophysical data. The part of
cosmoparticle physics, called cosmoarcheology, offers the set of
methods and tools probing such predictions.

Cosmoarcheology considers the results of observational cosmology
as the sample of the experimental data on the possible existence
and features of hypothetical phenomena predicted by particle
theory. To undertake the {\it Gedanken Experiment} with these
phenomena some theoretical framework to treat their origin and
evolution in the Universe should be assumed. As it was pointed out
in Ref. \citen{Cosmoarcheology} the choice of such framework is a
nontrivial problem in the modern cosmology.

Indeed, in the old Big Bang scenario any new phenomenon, predicted
by particle theory was considered in the course of the thermal
history of the Universe, starting from Planck times. The problem
is that the bedrock of the modern cosmology, namely, inflation,
baryosynthesis and dark matter, is also based on experimentally
unproven part of particle theory, so that the test for possible
effects of new physics implies the necessity to choose
the physical basis for such test. There are two possible solutions
for this problem: \begin{itemize}
                    \item a) crude model independent comparison of the
predicted effect with the observational data and
                    \item b) model
dependent treatment of considered effect, provided that the model,
predicting it, contains physical mechanism of inflation,
baryosynthesis and dark matter.
                  \end{itemize}

The basis for the approach (a) is that whatever happened in the
early Universe its results should not contradict the observed
properties of the modern Universe. The set of observational data
and, especially, the light element abundance and thermal spectrum
of microwave background radiation put severe constraint on the
deviation from thermal evolution after 1 s of expansion, what
strengthens the model independent conjectures of approach (a).

One can specify the new phenomena by their net contribution into
the cosmological density and by forms of their possible influence
on parameters of matter and radiation. In the first aspect we can
consider strong and weak phenomena. Strong phenomena can put
dominant contribution into the density of the Universe, thus
defining the dynamics of expansion in that period, whereas the
contribution of weak phenomena into the total density is always
subdominant. The phenomena are time dependent, being characterized
by their time-scale, so that permanent (stable) and temporary
(unstable) phenomena can take place. They can have homogeneous and
inhomogeneous distribution in space. The amplitude of density
fluctuations $\delta \equiv \delta \varrho/\varrho$ measures the
level of inhomogeneity relative to the total density, $\varrho$.
The partial amplitude $\delta_{i} \equiv \delta
\varrho_{i}/\varrho_{i}$ measures the level of fluctuations within
a particular component with density $\varrho_{i}$, contributing
into the total density $\varrho = \sum_{i} \varrho_{i}$. The case
$\delta_{i} \ge 1$ within the considered $i$-th component
corresponds to its strong inhomogeneity. Strong inhomogeneity is
compatible with the smallness of total density fluctuations, if
the contribution of inhomogeneous component into the total density
is small: $\varrho_{i} \ll \varrho$, so that $\delta \ll 1$ (see for review Ref. \citen{PBHrev}).

The phenomena can influence the properties of matter and radiation
either indirectly, say, changing of the cosmological equation of
state, or via direct interaction with matter and radiation. In the
first case only strong phenomena are relevant, in the second case
even weak phenomena are accessible to observational data. The
detailed analysis of sensitivity of cosmological data to various
phenomena of new physics are presented in Ref. \citen{book}.

The basis for the approach (b) is provided by a particle model, in
which inflation, baryosynthesis and nonbaryonic dark matter is
reproduced. Any realization of such physically complete basis for
models of the modern cosmology contains with necessity additional
model dependent predictions, accessible to cosmoarcheological
means. Here the scenario should contain all the details, specific
to the considered model, and the confrontation with the
observational data should be undertaken in its framework. In this
approach complete cosmoparticle physics models may be realized,
where all the parameters of particle model can be fixed from the
set of astrophysical, cosmological and physical constraints. Even
the details, related to cosmologically irrelevant predictions,
such as the parameters of unstable particles, can find the
cosmologically important meaning in these models. So, in the model
of horizontal unification \cite{Berezhiani1,Berezhiani2,Berezhiani3,Sakharov1},
the top quark or B-meson physics fixes the
parameters, describing the dark matter, forming the large scale
structure of the Universe, while in supersymmetric models experimental searches for
unstable SUSY particles fix
the parameters of SUSY dark matter candidates \cite{jungman}.

\subsection{Cosmophenomenology of new physics}
 \label{Cosmophenomenology}

To study the imprints of new physics in astrophysical data
cosmoarcheology implies the forms and means in which new physics
leaves such imprints. So, the important tool of cosmoarcheology in
linking the cosmological predictions of particle theory to
observational data is the {\it Cosmophenomenology} of new physics.
It studies the possible hypothetical forms of new physics, which
may appear as cosmological consequences of particle theory, and
their properties, which can result in observable effects.
\subsubsection{Stable relics. Freezing out. Charge symmetric case}
 \label{WIMPs}
The simplest primordial form of new physics is the gas of new
stable massive particles, originated from early Universe. For
particles with the mass $m$, at high temperature $T>m$ the
equilibrium condition, $$n \cdot \sigma v \cdot t > 1$$ is valid, if
their annihilation cross section $\sigma > 1/(m m_{Pl})$ is
sufficiently large to establish the equilibrium. At $T<m$ such
particles go out of equilibrium and their relative concentration
freezes out. This
is the main idea of calculation of primordial abundance for
Weakly Interacting Massive Particles (WIMPs, see e.g. Refs.
\citen{book,book3,Cosmoarcheology} for details).

If ordinary particles are among the products of WIMP annihilation, even their small fraction can annihilate in the Galaxy causing significant effect in cosmic rays and gamma background. This effect, first revealed in Ref. \citen{ZKKC} and then proved for even subdominant fraction of annihilating dark matter in Ref. \citen{DKKM}, is now in the basis of indirect dark matter searches in cosmic rays\cite{jungman}.

The process of WIMP annihilation to ordinary particles, considered in $t$channel,
determines their scattering cross section on ordinary particles and thus
relates the primordial abundance of WIMPs to their scattering rate in the
ordinary matter. Forming nonluminous massive halo of our Galaxy, WIMPs can penetrate
the terrestrial matter and scatter on nuclei in underground detectors. The strategy of
direct WIMP searches implies detection of recoil nuclei from this scattering.

The process inverse to annihilation of WIMPs corresponds to their production in collisions
of ordinary particles. It should lead to effects of missing mass and energy-momentum,
being the challenge for experimental search for production of dark matter candidates at accelerators,
e.g. at LHC.

\subsubsection{Stable relics. Decoupling}

More weakly interacting and/or more light species decouple from plasma
and radiation being relativistic
at $T \gg m$, when $$n \cdot \sigma v \cdot t \sim 1,$$
i.e. at $$T_{dec} \sim (\sigma m_{Pl})^{-1} \gg m.$$ After decoupling these species retain
their equilibrium distribution until they become non-relativistic at $T < m$.
Conservation of partial entropy in the cosmological expansion links the modern abundance
of these species to number density of relic photons with the account for the increase of
the photon number density due to the contribution of heavier ordinary particles, which were
in equilibrium in the period of decoupling.

For example, primordial neutrino decouple in the period, when relativistic electron-positron
plasma was present in the equilibrium. The account for increase of the number density of relic photons
due to electron-positron annihilation at $T<m_e$, where $m_e$ is the mass of electron, results in the
well known prediction of the Big Bang cosmology\cite{Weinberg,ZNSEU} $$n_{\nu{\bar\nu}}=\frac{3}{11}n_{\gamma},$$ where $n_{\nu{\bar\nu}}$ is the modern number density of a one species of primordial left-handed neutrinos (and the corresponding antineutrinos) and $n_{\gamma}=400 \cm^{-3}$ is the number density of CMB photons at the modern CMB temperature $T=2.7 \K$. Multiplying the predicted modern concentration of neutrinos by their mass, we obtain their contribution into the total density. This contribution should not exceed the total density, what gave early cosmological upper limits on neutrino mass. For the long time, it seemed possible that relic neutrinos can be the dominant form of cosmological dark matter and the corresponding neutrino-dominated Universe was considered as physical ground of Hot Dark Matter scenario of Large scale structure formation.  Experimental discovery of neutrino oscillations together with stringent upper limits on the mass of electron neutrino exclude this possibility. Moreover, even neutrino masses in the range of 1eV lead to features in the spectrum of density fluctuations that are excluded by the observational data of CMB.

Right handed neutrinos and left handed antineutrinos, involved in the seesaw mechanism of neutrino mass generation, are sterile relative to ordinary weak interaction. If these species were in thermal equilibrium in the early Universe, they should decouple much earlier, than ordinary neutrinos, in the period, when there were much more particle species (leptons, quarks, gluons,...) in the equilibrium, what leads to the primordial abundance of sterile neutrinos much smaller, than of the ordinary ones. Therefore cosmological constraints admit sterile neutrinos with the mass in the keV range. We refer to the Ref. \citen{keVnu} for the recent review of models of sterile neutrinos and their possible effects.

\subsubsection{Stable relics. SuperWIMPs}
The maximal
temperature, which is reached in inflationary Universe, is the
reheating temperature, $T_{r}$, after inflation. So, the very
weakly interacting particles with the annihilation cross section
$$\sigma < 1/(T_{r} m_{Pl}),$$ as well as very heavy particles with
the mass $$m \gg T_{r}$$ can not be in thermal equilibrium, and the
detailed mechanism of their production should be considered to
calculate their primordial abundance.

In particular, thermal production of gravitino in very early Universe is proportional to the reheating temperature $T_{r}$, what puts upper limit on this temperature from constraints on primordial gravitino abundance\cite{khlopovlinde,khlopovlinde2,khlopovlinde3,khlopov3,khlopov31,Karsten,Kawasaki}.
\subsubsection{Self interacting dark matter}\label{mirror}
Extensive hidden sector of particle theory can provide the existence of new interactions, which only new particles possess. Historically one of the first examples of such self-interacting dark matter was presented by the model of mirror matter. Mirror particles, first proposed by T. D. Lee and C. N. Yang in Ref. \citen{LeeYang} to restore equivalence of left- and right-handed co-ordinate systems in the presence of P- and C- violation in weak interactions, should be strictly symmetric by their properties to their ordinary twins. After discovery of CP-violation it was shown by I. Yu. Kobzarev, L. B. Okun and I. Ya. Pomeranchuk in Ref. \citen{KOP} that mirror partners cannot be associated with antiparticles and should represent a new set of symmetric partners for ordinary quarks and leptons with their own strong, electromagnetic and weak mirror interactions. It means that there should exist mirror quarks, bound in mirror nucleons by mirror QCD forces and mirror atoms, in which mirror nuclei are bound with mirror electrons by mirror electromagnetic interaction \cite{ZKrev,FootVolkas}. If gravity is the only common interaction for ordinary and mirror particles, mirror matter can be present in the Universe in the form of elusive mirror objects, having symmetric properties with ordinary astronomical objects (gas, plasma, stars, planets...), but causing only gravitational effects on the ordinary matter \cite{Blin1,Blin2}.

Even in the absence of any other common interaction except for gravity, the observational data on primordial helium abundance and upper limits on the local dark matter seem to exclude mirror matter, evolving in the Universe in a fully symmetric way in parallel with the ordinary baryonic matter\cite{Carlson,FootVolkasBBN}. The symmetry in cosmological evolution of mirror matter can be broken either by initial conditions\cite{zurabCV,zurab}, or by breaking mirror symmetry in the sets of particles and their interactions as it takes place in the shadow world\cite{shadow,shadow2}, arising in the heterotic string model. We refer to Refs.
\citen{newBook,OkunRev,Paolo} for current review of mirror matter and its cosmology.

If new particles possess new $y$-charge, interacting with massless bosons or intermediate bosons with sufficiently small mass ($y$-interaction),  for slow $y$-charged particles Coulomb-like factor
of "Gamov-Sommerfeld-Sakharov enhancement" \cite{Som,Sak,Sakhenhance} should be added in
the annihilation cross section
$$C_y=\frac{2 \pi \alpha_y/v}{1 - \exp{(-2 \pi \alpha_y/v)}},$$
where $v$ is relative velocity and $\alpha_y$ is the running gauge constant of $y$-interaction. This factor may not be essential in the period of particle freezing out in the early Universe (when $v$ was only few times smaller than $c$), but can cause strong enhancement in the effect of annihilation of nonrelativistic dark matter particles in the Galaxy.
\subsubsection{Subdominant dark matter}
If charge symmetric stable particles (and their antiparticles) represent
only subdominant fraction of the cosmological dark matter, more detailed analysis
of their distribution in space, of their condensation in galaxies,
of their capture by stars, Sun and Earth, as well as effects of
their interaction with matter and of their annihilation provides
more sensitive probes for their existence.

In particular,
hypothetical stable neutrinos of 4th generation with mass about 50
GeV should be the subdominant form of modern dark
matter, contributing less than 0,1 \% to the total density
\cite{ZKKC,DKKM}. However, direct experimental search for cosmic
fluxes of weakly interacting massive particles (WIMPs) may be
sensitive to existence of such component (see Refs.
\citen{DAMA,DAMA-review,Bernabei:2008yi,DAMARev,CDMS,CDMS2,CDMS3,xenon,cogent} and references therein). It was
shown in Refs. \citen{Fargion99,Grossi,Belotsky,Belotsky2} that
annihilation of 4th neutrinos and their antineutrinos in the Galaxy
is severely constrained by the measurements of gamma-background, cosmic positrons
and antiprotons. 4th neutrino
annihilation inside the Earth should lead to the flux of underground
monochromatic neutrinos of known types, which can be traced in the
analysis of the already existing and future data of underground
neutrino detectors \cite{Belotsky,BKS1,BKS2,BKS3}.
\subsubsection{Decaying dark matter}
Decaying particles with lifetime $\tau$, exceeding the age of the
Universe, $t_{U}$, $\tau > t_{U}$, can be treated as stable. By
definition, primordial stable particles survive to the present time
and should be present in the modern Universe. The net effect of
their existence is given by their contribution into the total
cosmological density. However, even small effect of their decay
can lead to significant contribution to cosmic rays and gamma background\cite{ddm}.
Leptonic decays of dark matter are considered as possible explanation of
the cosmic positron excess, measured in the range above 10 GeV by PAMELA\cite{pamela}, FERMI/LAT\cite{lat} and AMS02\cite{ams2}
(see Ref. \citen{amsRev} for the review of AMS02 experiment).
\subsubsection{Charge asymmetry of dark matter}
The fact that particles are not absolutely stable means that the corresponding charge is not strictly conserved and generation particle charge asymmetry is possible, as it is assumed for ordinary baryonic matter. At sufficiently strong particle annihilation cross section excessive particles (antiparticles) can dominate in the relic density, leaving exponentially small admixture of their antiparticles (particles) in the same way as primordial excessive baryons dominate over antibaryons in baryon asymmetric Universe. In this case {\it Asymmetric dark matter} doesn't lead to significant effect of particle annihilation in the modern Universe and can be searched for either directly in underground detectors or indirectly by effects of decay or condensation and structural transformations of e.g. neutron stars (see Ref. \citen{adm} for recent review and references). If particle annihilation isn't strong enough, primordial pairs of particles and antiparticles dominate over excessive particles (or antiparticles) and this case has no principle difference from the charge symmetric case. In particular, for very heavy charged leptons (with the mass above 1 TeV), like "tera electrons"\cite{Glashow}, discussed in \ref{asymmetry}, their annihilation due to electromagnetic interaction is too weak to provide effective suppression of primordial tera electron-positron pairs relative to primordial asymmetric excess\cite{BKSR1}.
\subsubsection{Charged stable relics. Dark atoms}
New particles with electric charge and/or strong interaction can
form anomalous atoms and contain in the ordinary matter as anomalous
isotopes. For example, if the lightest quark of 4th generation is
stable, it can form stable charged hadrons, serving as nuclei of
anomalous atoms of e.g. anomalous helium
\cite{BKSR1,BKS,BKSR,FKS,I,BKSR4}. Therefore, stringent upper limits on anomalous isotopes, especially, on anomalous hydrogen put severe constraints on the existence of new stable charged particles. However, as we discuss in Section \ref{flavor}, stable doubly charged particles can not only exist, but even dominate in the cosmological dark matter, being effectively hidden in neutral "dark atoms"\cite{DADM}.
\subsubsection{Unstable particles}
 \label{unstable}
Primordial unstable particles with the lifetime, less than the age
of the Universe, $\tau < t_{U}$, can not survive to the present
time. But, if their lifetime is sufficiently large to satisfy the
condition $\tau \gg (m_{Pl}/m) \cdot (1/m)$, their existence in
early Universe can lead to direct or indirect traces\cite{khlopov7}.

Weakly interacting particles, decaying to invisible modes, can influence Large Scale Structure formation.
Such decays prevent formation of the structure, if they take place before the structure is formed.
Invisible products of decays after the structure is formed should contribute in the cosmological dark energy.
The Unstable Dark matter scenarios\cite{Sakharov1,UDM,UDM1,UDM2,UDM3,berezhiani4,berezhiani5,TSK,GSV} implied weakly interacting particles that form the structure on the matter dominated stage and then decay to invisible modes after the structure is formed.

Cosmological
flux of decay products contributing into the cosmic and gamma ray
backgrounds represents the direct trace of unstable particles\cite{khlopov7,sedelnikov}. If
the decay products do not survive to the present time their
interaction with matter and radiation can cause indirect trace in
the light element abundance\cite{khlopovlinde3,khlopov3,khlopov31,DES} or in the fluctuations of thermal
radiation\cite{UDM4}.

If the particle lifetime is much less than $1$s the
multi-step indirect traces are possible, provided that particles
dominate in the Universe before their decay. On the dust-like
stage of their dominance black hole formation takes place, and the
spectrum of such primordial black holes traces the particle
properties (mass, frozen concentration, lifetime) \cite{polnarev,khlopov0,polnarev0}.
The particle decay in the end of dust like stage influences the
baryon asymmetry of the Universe. In any way cosmophenomenoLOGICAL
chains link the predicted properties of even unstable new
particles to the effects accessible in astronomical observations.
Such effects may be important in the analysis of the observational
data.
\subsubsection{Phase transitions}
Parameters of new stable and metastable particles are also
determined by a pattern of particle symmetry breaking. This pattern
is reflected in a succession of phase transitions in the early
Universe. First order phase transitions proceed through bubble
nucleation, which can result in black hole formation (see e.g.
Refs. \citen{kkrs} and \citen{book2} for review and references). Phase
transitions of the second order can lead to formation of topological
defects, such as walls, string or monopoles. The observational data
put severe constraints on magnetic monopole \cite{kz} and cosmic
wall production \cite{okun}, as well as on the parameters of cosmic
strings \cite{zv1,zv2}. Structure of cosmological defects can be
changed in succession of phase transitions. More complicated forms
like walls-surrounded-by-strings can appear. Such structures can be
unstable, but their existence can leave a trace in nonhomogeneous
distribution of dark matter and give rise to large scale structures
of nonhomogeneous dark matter like {\it archioles}
\cite{Sakharov2,kss,kss2}. This effect should be taken into account in the analysis of
cosmological effects of weakly interacting slim particles (WISPs) (see Ref. \citen{jaeckel} for current review) that can play the role of cold dark matter in spite of their small mass.

\subsection{Structures from succession of U(1) phase transitions}\label{structures}
A wide class of particle models possesses a symmetry breaking
pattern, which can be effectively described by
pseudo-Nambu--Goldstone (PNG) field and which corresponds to
formation of unstable topological defect structure in the early
Universe (see Ref. \citen{book2} for review and references). The
Nambu--Goldstone nature in such an effective description reflects
the spontaneous breaking of global U(1) symmetry, resulting in
continuous degeneracy of vacua. The explicit symmetry breaking at
smaller energy scale changes this continuous degeneracy by discrete
vacuum degeneracy. The character of formed structures is  different
for phase transitions, taking place on post-inflationary and
inflationary stages.
\subsubsection{Large scale correlations of axion field}\label{axion}
At high temperatures such a symmetry breaking pattern implies the
succession of second order phase transitions. In the first
transition, continuous degeneracy of vacua leads, at scales
exceeding the correlation length, to the formation of topological
defects in the form of a string network; in the second phase
transition, continuous transitions in space between degenerated
vacua form surfaces: domain walls surrounded by strings. This last
structure is unstable, but, as was shown in the example of the
invisible axion \cite{Sakharov2,kss,kss2}, it is reflected in the
large scale inhomogeneity of distribution of energy density of
coherent PNG (axion) field oscillations. This energy density is
proportional to the initial value of phase, which acquires dynamical
meaning of amplitude of axion field, when axion mass is switched on
in the result of the second phase transition.

The value of phase changes by $2 \pi$ around string. This strong
nonhomogeneity of phase leads to corresponding nonhomogeneity of
energy density of coherent PNG (axion) field oscillations. Usual
argument (see e.g. Ref. \citen{kim} and references therein) is essential
only on scales, corresponduing to mean distance between strings.
This distance is small, being of the order of the scale of
cosmological horizon in the period, when PNG field oscillations
start. However, since the nonhomogeneity of phase follows the
pattern of axion string network this argument misses large scale
correlations in the distribution of oscillations' energy density.

Indeed, numerical analysis of string network (see review in the
Ref. \citen{vs}) indicates that large string loops are strongly suppressed
and the fraction of about 80\% of string length, corresponding to
long loops, remains virtually the same in all large scales. This
property is the other side of the well known scale invariant
character of string network. Therefore the correlations of energy
density should persist on large scales, as it was revealed in Refs.
\citen{Sakharov2,kss,kss2}.

The large
scale correlations in topological defects and their imprints in
primordial inhomogeneities is the indirect effect of inflation, if
phase transitions take place after reheating of the Universe.
Inflation provides in this case the equal conditions of phase
transition, taking place in causally disconnected regions.
\subsubsection{Primordial seeds for Active Galactic Nuclei}\label{AGN}
If the phase transitions take place on inflational stage new forms
of primordial large scale correlations appear. The example of
global U(1) symmetry, broken spontaneously in the period of
inflation and successively broken explicitly after reheating, was
considered in Ref. \citen{RKS}. In this model, spontaneous U(1)
symmetry breaking at inflational stage is induced by the vacuum
expectation value $\langle \psi \rangle = f$ of a complex scalar
field $\Psi = \psi \exp{(i \theta)}$, having also explicit
symmetry breaking term in its potential $V_{eb} = \Lambda^{4} (1 -
\cos{\theta})$. The latter is negligible in the period of
inflation, if $f \gg \Lambda$, so that there appears a valley
relative to values of phase in the field potential in this period.
Fluctuations of the phase $\theta$ along this valley, being of the
order of $\Delta \theta \sim H/(2\pi f)$ (here $H$ is the Hubble
parameter at inflational stage) change in the course of inflation
its initial value within the regions of smaller size. Owing to
such fluctuations, for the fixed value of $\theta_{60}$ in the
period of inflation with {\it e-folding} $N=60$ corresponding to
the part of the Universe within the modern cosmological horizon,
strong deviations from this value appear at smaller scales,
corresponding to later periods of inflation with $N < 60$. If
$\theta_{60} < \pi$, the fluctuations can move the value of
$\theta_{N}$ to $\theta_{N} > \pi$ in some regions of the
Universe. After reheating, when the Universe cools down to
temperature $T = \Lambda$ the phase transition to the true vacuum
states, corresponding to the minima of $V_{eb}$ takes place. For
$\theta_{N} < \pi$ the minimum of $V_{eb}$ is reached at
$\theta_{vac} = 0$, whereas in the regions with $\theta_{N} > \pi$
the true vacuum state corresponds to $\theta_{vac} = 2\pi$. For
$\theta_{60} < \pi$ in the bulk of the volume within the modern
cosmological horizon $\theta_{vac} = 0$. However, within this
volume there appear regions with $\theta_{vac} = 2\pi$. These
regions are surrounded by massive domain walls, formed at the
border between the two vacua. Since regions with $\theta_{vac} =
2\pi$ are confined, the domain walls are closed. After their size
equals the horizon, closed walls can collapse into black holes.
The minimal mass of such black hole is determined by the condition
that it's Schwarzschild radius, $r_{g} = 2 G M/c^{2}$ exceeds the
width of the wall, $l \sim f/\Lambda^{2}$, and it is given by
$M_{min} \sim f (m_{Pl}/\Lambda)^{2}$. The maximal mass is
determined by the mass of the wall, corresponding to the earliest
region $\theta_{N} > \pi$, appeared at inflational stage.

  This mechanism can lead to formation
of primordial black holes of a whatever large mass (up to the mass
of AGNs \cite{AGN,DER1}, see for latest review Ref. \citen{PBHrev}). Such black
holes appear in the form of primordial black hole clusters,
exhibiting fractal distribution in space
\cite{KRS,Khlopov:2004sc,book2}. It can shed new light on the
problem of galaxy formation \cite{book2,DER1}.
\begin{figure}
\begin{center}
\includegraphics[scale=0.7]{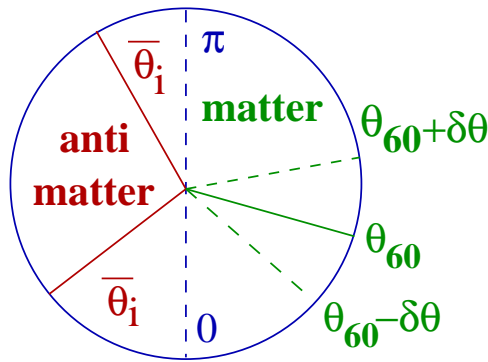}
\caption{ The inflational evolution of the
phase (taken from the Ref. \citen{ere}). The phase  $\theta_{60}$ sits in the range $[\pi ,0 ]$ at the
beginning of inflation and makes Brownian step
$\delta\theta_{eff}=H_{infl}/(2\pi f_{eff})$  at each e--fold. The
typical wavelength of the fluctuation $\delta\theta$ is equal to
$H^{-1}_{infl}$. The whole domain  $H^{-1}_{infl}$, containing phase
$\theta_{N}$ gets  divided, after one e--fold, into  $e^3$ causally
disconnected domains of radius $H^{-1}_{infl}$. Each new domain
contains almost homogeneous phase value
$\theta_{N-1}=\theta_{N}\pm\delta\theta_{eff}$. Every successive
e-fold this process repeats in every domain.}
\label{pnginfl}
\end{center}
\end{figure}
\subsubsection{Antimatter in Baryon asymmetric Universe?}\label{antimatter}
Primordial strong inhomogeneities can also appear in the baryon
charge distribution. The appearance of antibaryon domains in the
baryon asymmetrical Universe, reflecting the inhomogeneity of
baryosynthesis, is the profound signature of such strong
inhomogeneity \cite{CKSZ}. On the example of the model of
spontaneous baryosynthesis (see Ref. \citen{Dolgov} for review) the
possibility for existence of antimatter domains, surviving to the
present time in inflationary Universe with inhomogeneous
baryosynthesis was revealed in \cite{KRS2}.

The mechanism of
spontaneous baryogenesis \cite{Dolgov} implies the existence of a
complex scalar field $\chi =(f/\sqrt{2})\exp{(\theta )}$ carrying
the baryonic charge. The $U(1)$ symmetry, which corresponds to the
baryon charge, is broken spontaneously and explicitly. The explicit
breakdown of $U(1)$ symmetry is caused by the phase-dependent term
 \beq\label{expl} V(\theta )=\Lambda^4(1-\cos\theta ).
 \eeq
The possible baryon and lepton number violating interaction
of the field $\chi$ with matter fields can have the following
structure \cite{Dolgov} \beq\label{leptnumb} {\cal
L}=g\chi\bar QL+{\rm h.c.}, \eeq where fields $Q$ and $L$ represent
a heavy quark and lepton, coupled to the ordinary matter fields.

In the early Universe, at a time when the friction term, induced by
the Hubble constant, becomes comparable with the angular mass
$m_{\theta}=\frac{\Lambda^2}{f}$, the phase $\theta$ starts to
oscillate around the minima of the PNG potential and decays into
matter fields according to (\ref{leptnumb}). The coupling
(\ref{leptnumb}) gives rise to the following \cite{Dolgov}: as
the phase starts to roll down in the clockwise direction (Fig.~\ref{pnginfl}),  it preferentially creates
excess of baryons over antibaryons, while the opposite is true as it
starts to roll down in the opposite direction.

The fate of such antimatter regions depends on their size. If the
physical size of some of them is larger than  the critical surviving size
$L_c=8h^2$ kpc~\cite{KRS2}, they survive annihilation with surrounding matter.
Evolution of
sufficiently dense antimatter domains can lead to formation of
antimatter globular clusters \cite{GC}. The existence of such
cluster in the halo of our Galaxy should lead to the pollution of
the galactic halo by antiprotons. Their annihilation can reproduce
\cite{Golubkov} the observed galactic gamma background in the
range tens-hundreds MeV. The prediction of antihelium component of
cosmic rays \cite{ANTIHE}, accessible to future searches for
cosmic ray antinuclei in PAMELA and AMS II experiments, as well as
of antimatter meteorites \cite{ANTIME} provides the direct
experimental test for this hypothesis.

So the primordial strong inhomogeneities in the distribution of
total, dark matter and baryon density in the Universe is the new
important phenomenon of cosmological models, based on particle
models with hierarchy of symmetry breaking.

\section{Primordial Black Holes as cosmological reflection of particle structure}
 \label{pbhs}
It was probably Pierre-Simon Laplace \cite{Laplace} in the beginning
of XIX century, who noted first that in very massive stars escape
velocity can exceed the speed of light and light can not come from
such stars. This conclusion made in the framework of Newton
mechanics and Newton corpuscular theory of light has further
transformed into the notion of "black hole" in the framework of
general relativity and electromagnetic theory. Any object of mass
$M$ can become a black hole, being put within its gravitational
radius $r_g=2 G M/c^2.$ At present time black holes (BH) can be
created only by a gravitational collapse of compact objects with
mass more than about three Solar mass \cite{1,ZNRA}. It can be a
natural end of massive stars or can result from evolution of dense
stellar clusters. However in the early Universe there were no limits
on the mass of BH. Ya.B. Zeldovich and I.D. Novikov (see Ref. \citen{ZN})
noticed that if cosmological expansion stops in some region, black
hole can be formed in this region within the cosmological horizon.
It corresponds to strong deviation from general expansion and
reflects strong inhomogeneity in the early Universe.  There are
several mechanisms for such strong inhomogeneity and we'll trace
their links to cosmological consequences of particle theory.

Primordial Black Holes (PBHs) are a very sensitive cosmological
probe for physics phenomena occurring in the early Universe. They
could be formed by many different mechanisms, {\it e.g.}, initial
density inhomogeneities \cite{hawking1,hawkingCarr} and non-linear
metric perturbations
\cite{Bullock:1996at,Ivanov:1997ia}, blue spectra of
density fluctuations
\cite{Khlopov:1984wc,polnarev,Lidsey:1995ir,Kotok:1998rp,
Dubrovich02,Sendouda:2006nu}, a softening of the equation of state
\cite{canuto,Khlopov:1984wc,polnarev}, development of gravitational
instability on early dust-like stages of dominance of supermassive
particles and scalar fields
\cite{khlopov0,polnarev0,polnarev1,khlopov1} and evolution of
gravitationally bound objects formed at these stages
\cite{Kalashnikov,Kadnikov}, collapse of cosmic strings
\cite{hawking2,Polnarev:1988dh,Hansen:2000jv,Cheng:1996du,Nagasawa:2005hv}
and necklaces \cite{Matsuda:2005ez}, a double inflation scenario
\cite{nas,Kim:1999xg,Yamaguchi:2001zh}, first order
phase transitions \cite{hawking3,pt1Jedamzik,kkrs,kkrs1,kkrs2}, a
step in the power spectrum \cite{Sakharov0,polarski1}, etc. (see Refs.
\citen{polnarev,book,book3,Carr:2003bj,book2} for a review).

Being formed, PBHs should retain in the Universe and, if survive to
the present time, represent a specific form of dark matter
\cite{khlopov7,Ivanov:1994pa,book,book3,Blais:2002nd,Chavda:2002cj,
Afshordi:2003zb,book2,Chen:2004ft}. Effect of PBH evaporation by
S.W.Hawking \cite{hawking4} makes evaporating PBHs a source of
fluxes of products of evaporation, particularly of $\gamma$
radiation \cite{Page:1976wx}. MiniPBHs with mass below $10^{14}$~g
evaporate completely and do not survive to the present time.
However, effect of their evaporation should cause influence on
physical processes in the early Universe, thus providing a test for
their existence by methods of cosmoarcheology
\cite{Cosmoarcheology}, studying cosmological imprints of new
physics in astrophysical data. In a wide range of parameters the
predicted effect of PBHs contradicts the data and it puts
restrictions on mechanism of PBH formation and the underlying
physics of very early Universe. On the other hand, at some fixed
values of parameters, PBHs or effects of their evaporation can
provide a nontrivial solution for astrophysical problems.

Various aspects of PBH physics, mechanisms of their formation,
evolution and effects are discussed in Refs.
\citen{carr1,carrMG,LGreen,khlopov6,polnarev,Grillo:1980uj,
Chapline:1975tn,Hayward:1989jq,Yokoyama:1995ex,Kim:1996hr,
Heckler:1997jv,green,Niemeyer:1997mt,Kribs:1999bs,
Green:2004wb,Yokoyama:1998pt,Yokoyama:1998xd,Yokoyama:1999xi,
Bringmann:2001yp,Dimopoulos:2003ce,Nozari:2007kv,LythMalik,
Harada:2004pf,
Custodio:2005en,Bousso:1995cc,Bousso:1996wy,Elizalde:1999dw,
Nojiri:1999vv,Bousso:1999iq,Silk:2000em,Polarski:2001jk,
Barrow:1996jk,Paul:2000jb,Paul:2001yt,Paul:2005bk,Polarski:2001yn,
Carr:1993aq,Yokoyama:1998qw,Kaloper:2004yj,
Stojkovic:2005zh,Ahn2011} particularly specifying PBH
formation and effects in braneworld cosmology
\cite{Guedens:2002km,Guedens:2002sd,Clancy:2003zd,
Tikhomirov:2005bt}, on inflationary preheating
\cite{Bassett:2000ha}, formation of PBHs in QCD phase transition
\cite{Jedamzik:1998hc}, properties of superhorizon
BHs \cite{Harada:2005sc,Harada:2006gn}, role of PBHs in
baryosynthesis
\cite{Grillo:1980rt,Barrow:1990he,Turner:1979bt,Upadhyay:1999vk,
Bugaev:2001xr}, effects of PBH evaporation in the early Universe and
in modern cosmic ray, neutrino and gamma fluxes
\cite{mujana,Fegan:1978zn,Green:2001kw,Frampton:2005fk,
MacGibbon:1990zk,MacGibbon:1991tj,Halzen:1991uw,Halzen:1995hu,
Bugaev:2000bz,Bugaev:2002yt,Volkova:1994fb,Gibilisco:1996ft,
Golubkov:2000qy,He:2002vz,Gibilisco:1996dk,Custodio:2002jv,
Sendouda:2003dc,Maki:1995pa,barraupbar,Barrau:2002mc,
Wells:1998jv,Cline:1996uk,Xu:1998hn,Cline:1998fx,
Barrau:1999sk,Derishev:1999xn,Tikhomirov:2004rs,barrau,
barrauprd}, in creation of hypothetical particles
\cite{Bell:1998jk,lemoine,green1,barrau2}, PBH clustering and
creation of supermassive BHs
\cite{DER1,Bean:2002kx,Duechting:2004dk,Chisholm:2005vm}, effects in cosmic rays
and colliders from PBHs in low scale gravity models
\cite{barrauADDBH,barrauADDac}. Here we outline the role of PBHs as
a link in cosmoarcheoLOGICAL chain, connecting cosmological
predictions of particle theory with observational data. We discuss
the way, in which spectrum of PBHs reflects properties of superheavy
metastable particles and of phase transitions on inflationary and
post-inflationary stages. We
illustrate in subsection \ref{dust} some mechanisms of PBH formation on
stage of dominance of superheavy particles and fields (subsubsection
\ref{particles}) and from second order phase transition on
inflationary stage. Effective mechanism of BH formation during
bubble nucleation provides a sensitive tool to probe existence of
cosmological first order phase transitions by PBHs (subsection
\ref{phasetransitions}). Existence of stable remnants of PBH
evaporation can strongly increase the sensitivity of such probe and
we demonstrate this possibility in subsection \ref{gravitino} on an
example of gravitino production in PBH evaporation. Being formed
within cosmological horizon, PBHs seem to have masses much less than
the mass of stars, constrained by small size of horizon in very
early Universe. However, if phase transition takes place on
inflationary stage, closed walls of practically any size can be
formed (subsubsection \ref{walls}) and their successive collapse can
give rise to clouds of massive black holes, which can play the role
of seeds for galaxies (subsection \ref{MBHwalls}).


\subsection{PBHs from early dust-like stages}\label{dust}
A possibility to form a black hole is highly improbable in
homogeneous expanding Universe, since it implies metric fluctuations
of order 1. For metric fluctuations distributed according to
Gaussian law with dispersion \begin{equation}
\label{DispBH}\left\langle \delta^2 \right\rangle \ll
1\end{equation} a probability for fluctuation of order 1 is
determined by exponentially small tail of high amplitude part of
this distribution. This probability can be even more suppressed in a
case of non-Gaussian fluctuations \cite{Bullock:1996at}.

In the Universe with equation of state \begin{equation}
\label{EqState}p=\gamma \epsilon,\end{equation} with numerical
factor $\gamma$ being in the range \begin{equation}
\label{FacState}0 \le \gamma \le 1\end{equation} a probability to
form black hole from fluctuation within cosmological horizon is
given by (see e.g. \cite{book,book3} for review and references)
\begin{equation}
\label{ProbBH}W_{PBH} \propto \exp \left(-\frac{\gamma^2}{2
\left\langle \delta^2 \right\rangle}\right).
\end{equation}
It provides exponential sensitivity of PBH spectrum to softening of
equation of state in early Universe ($\gamma \rightarrow 0$) or to
increase of ultraviolet part of spectrum of density fluctuations
($\left\langle \delta^2 \right\rangle \rightarrow 1$). These
phenomena can appear as cosmological consequence of particle theory.
\subsubsection{Dominance of superheavy particles in early Universe}\label{particles}
Superheavy particles can not be studied at accelerators directly. If
they are stable, their existence can be probed by cosmological
tests, but there is no direct link between astrophysical data and
existence of superheavy metastable particles with lifetime $\tau \ll
1s$. It was first noticed in Ref. \citen{khlopov0} that dominance of such
particles in the Universe before their decay at $t \le \tau$ can
result in formation of PBHs, retaining in Universe after the
particles decay and keeping some information on particle properties
in their spectrum. It provided though indirect but still a
possibility to probe existence of such particles in astrophysical
observations. Even the absence of observational evidences for PBHs
is important. It puts restrictions on allowed properties of
superheavy metastable particles, which might form such PBHs on a
stage of particle dominance, and thus constrains parameters of
models, predicting these particles.

After reheating, at \begin{equation} \label{Eareq}T <
T_0=rm\end{equation} particles with mass $m$ and relative abundance
$r=n/n_r$ (where $n$ is frozen out concentration of particles and
$n_r$ is concentration of relativistic species) must dominate in the
Universe before their decay. Dominance of these nonrelativistic
particles at $t>t_0$, where
\begin{equation}
\label{EarMD}t_0=\frac{m_{pl}}{T_0^2},\end{equation} corresponds to
dust like stage with equation of state $p=0,$ at which particle
density fluctuations grow as\begin{equation}
\label{dens}\delta(t)=\frac{\delta \rho}{\rho} \propto t^{2/3}
\end{equation}
 and development of gravitational instability results in formation
of gravitationally bound systems, which decouple at \begin{equation}
\label{decoup}t \sim t_f \approx t_i
\delta(t_i)^{-3/2}\end{equation} from general cosmological
expansion, when $\delta(t_f)\sim 1$ for fluctuations, entering
horizon at $t=t_i>t_0$ with amplitude $\delta(t_i)$.

Formation of these systems can result in black home formation either
immediately after the system decouples from expansion or in result
of evolution of initially formed nonrelativistic gravitationally
bound system.
\subsubsection{Direct PBH formation}\label{particles}
If density fluctuation is especially homogeneous and isotropic, it
directly collapses to BH as soon as the amplitude of fluctuation
grows to 1 and the system decouples from expansion. A probability
for direct BH formation in collapse of such homogeneous and
isotropic configurations gives minimal estimation of BH formation on
dust-like stage.

This probability was calculated in Ref. \citen{khlopov0} with the use of
the following arguments. In the period $t \sim t_f$, when
fluctuation decouples from expansion, its configuration is defined
by averaged density $\rho_1$, size $r_1$, deviation from sphericity
$s$ and by inhomogeneity $u$ of internal density distribution within
the fluctuation.  Having decoupled from expansion, the configuration
contracts and the minimal size to which it can contract is
\begin{equation} \label{sphcontr}r_{min} \sim s r_1,\end{equation}
being determined by a deviation from sphericity
\begin{equation} \label{spheric}s=\max\{\left\vert\gamma_1-\gamma_2\right\vert,\left\vert\gamma_1-
\gamma_2\right\vert,\left\vert\gamma_1-\gamma_2\right\vert\},\end{equation}
where $\gamma_1$, $\gamma_2$ and $\gamma_3$ define a deformation of
configuration along its three main orthogonal axes. It was first
noticed in Ref. \citen{khlopov0} that to form a black hole in result of
such contraction it is sufficient that configuration returns to the
size \begin{equation} \label{rminBH}r_{min} \sim r_g \sim t_i \sim
\delta(t_i) r_1,\end{equation} which had the initial fluctuation
$\delta(t_i)$, when it entered horizon at cosmological time $t_i$.
If
\begin{equation} \label{spher}s \le \delta(t_i),\end{equation}
configuration is sufficiently isotropic to concentrate its mass in
the course of collapse within its gravitational radius, but such
concentration also implies sufficient homogeneity of configuration.
Density gradients can result in gradients of pressure, which can
prevent collapse to BH. This effect does not take place for
contracting collisionless gas of weakly interacting massive
particles, but due to inhomogeneity of collapse the particles, which
have already passed the caustics can free stream beyond the
gravitational radius, before the whole mass is concentrated within
it. Collapse of nearly spherically symmetric dust configuration is
described by Tolmen solution. It's analysis
\cite{polnarev0,polnarev1,KP,polnarev} has provided a constraint on
the inhomogeneity $u=\delta \rho_1/\rho_1$ within the configuration.
It was shown that both for collisionless and interacting particles
the condition \begin{equation}
\label{inhom}u<\delta(t_i)^{3/2}\end{equation} is sufficient for
configuration to contract within its gravitational radius.

A probability for direct BH formation is then determined by a
product of probability for sufficient initial sphericity $W_s$ and
homogeneity $W_u$ of configuration, which is determined by the phase
space for such configurations. In a calculation of $W_s$ one should
take into account that the condition (\ref{spher}) implies 5
conditions for independent components of tensor of deformation
before its diagonalization (2 conditions for three diagonal
components to be close to each other and 3 conditions for
nondiagonal components to be small). Therefore, the probability of
sufficient sphericity is given by
\cite{khlopov0,polnarev0,polnarev1,KP,polnarev} \begin{equation}
\label{Wspher}W_s \sim \delta(t_i)^{5}\end{equation} and together
with the probability for sufficient homogeneity \begin{equation}
\label{Winhom}W_u \sim \delta(t_i)^{3/2}\end{equation} results in
the strong power-law suppression of probability for direct BH
formation \beq \label{WPBH} W_{PBH} = W_s \cdot W_u \sim
\delta(t_i)^{13/2}.\eeq Though this calculation was originally done
in Refs. \citen{khlopov0,polnarev0,polnarev1,KP,polnarev} for Gaussian
distribution of fluctuations, it does not imply specific form of
high amplitude tail of this distribution and thus should not change
strongly in a case of non-Gaussian fluctuations
\cite{Bullock:1996at}.

The mechanism
\cite{khlopov0,polnarev0,polnarev1,KP,polnarev,book,book3} is
effective for formation of PBHs with mass in an interval \beq
\label{Mint}M_0 \le M \le M_{bhmax}.\eeq The minimal mass
corresponds to the mass within cosmological horizon in the period $t
\sim t_0,$ when particles start to dominate in the Universe and it
is equal to
\cite{khlopov0,polnarev0,polnarev1,KP,polnarev,book,book3}\beq
\label{MBHmin} M_{0} = \frac{4 \pi}{3} \rho t^3_0 \approx
m_{pl}(\frac{m_{pl}}{r m})^2.\eeq
 The maximal mass is indirectly determined by the condition
 \beq
\label{Mconmax}\tau = t(M_{bhmax}) \delta(M_{bhmax})^{-3/2}\eeq that
fluctuation
 in the considered scale $M_{bhmax}$, entering the horizon at $t(M_{bhmax})$
 with an amplitude $\delta(M_{bhmax})$ can manage to grow up to nonlinear stage,
 decouple and collapse before particles decay at $t=\tau.$
 For scale invariant spectrum $\delta(M)=\delta_0$ the maximal mass
 is given by
\cite{book2}\beq \label{MBHmax} M_{bhmax} = m_{pl}
\frac{\tau}{t_{Pl}} \delta_0^{-3/2} =m_{pl}^2 \tau
\delta_0^{-3/2}.\eeq The probability, given by Eq.(\ref{WPBH}), is
also appropriate for formation of PBHs on dust-like preheating stage
after inflation \cite{khlopov1,book,book3}. The simplest example of
such stage can be given with the use of a model of homogeneous
massive scalar field \cite{book,book3}. Slow rolling of the field in
the period $t \ll 1/m$ (where $m$ is the mass of field) provides
chaotic inflation scenario, while at $t
> 1/m$ the field oscillates with period $1/m$. Coherent oscillations
of the field correspond to an averaged over period of oscillations
dust-like equation of state $p=0,$ at which gravitational
instability can develop. The minimal mass in this case corresponds
to the Jeans mass of scalar field, while the maximal mass is also
determined by a condition that fluctuation grows and collapses
before the scalar field decays and reheats the Universe.

The probability $W_{PBH}(M)$ determines the fraction of total
density \beq \label{beta}\beta(M)=\frac{\rho_{PBH}(M)}{\rho_{tot}}
\approx W_{PBH}(M),\eeq corresponding to PBHs with mass $M$. For
$\delta(M) \ll 1$ this fraction, given by Eq.(\ref{WPBH}), is small.
It means that the bulk of particles do not collapse directly in
black holes, but form gravitationally bound systems. Evolution of
these systems can give much larger amount of PBHs, but it strongly
depends on particle properties.
\subsubsection{Evolutional formation of PBHs}\label{particles}
Superweakly interacting particles form gravitationally bound systems
of collisionless gas, which remind modern galaxies with
collisionless gas of stars. Such system can finally collapse to
black hole, but energy dissipation in it and consequently its
evolution is a relatively slow process \cite{ZPod,book,book3}. The
evolution of these systems is dominantly determined by evaporation
of particles, which gain velocities, exceeding the parabolic
velocity of system. In the case of binary collisions the evolution
timescale can be roughly estimated \cite{ZPod,book,book3} as \beq
\label{tevbin} t_{ev} = \frac{N}{\ln{N}} t_{ff}\eeq for
gravitationally bound system of $N$ particles, where the free fall
time $t_{ff}$ for system with density $\rho$ is $t_{ff} \approx (4
\pi G \rho)^{-1/2}.$ This time scale can be shorter due to
collective effects in collisionless gas \cite{GurSav} and be at
large $N$ of the order of \beq \label{tevcol} t_{ev} \sim N^{2/3}
t_{ff}.\eeq However, since the free fall time scale for
gravitationally bound systems of collisionless gas is of the order
of cosmological time $t_f$ for the period, when these systems are
formed, even in the latter case the particles should be very long
living $\tau \ll t_f$ to form black holes in such slow evolutional
process.

The evolutional time scale is much smaller for gravitationally bound
systems of superheavy particles, interacting with light relativistic
particles and radiation. Such systems have analogy with stars, in
which evolution time scale is defined by energy loss by radiation.
An example of such particles give superheavy color octet fermions of
asymptotically free SU(5) model\cite{Kalashnikov} or magnetic
monopoles of GUT models. Having decoupled from expansion, frozen out
particles and antiparticles can annihilate in gravitationally bound
systems, but detailed numerical simulation\cite{Kadnikov} has
shown that annihilation can not prevent collapse of the most of mass
and the timescale of collapse does not exceed the cosmological time
of the period, when the systems are formed.

\subsection{Spikes from phase transitions on inflationary stage}
Scale non-invariant spectrum of fluctuations, in which amplitude of
small scale fluctuations is enhanced, can be another factor,
increasing the probability of PBH formation. The simplest functional
form of such spectrum is represented by a blue spectrum with a power
law dispersion \beq\left\langle \delta^2(M) \right\rangle \propto
M^{-k},\eeq with amplitude of fluctuations growing at $k>0$ to small
$M$. The realistic account for existence of other scalar fields
together with inflaton in the period of inflation can give rise to
spectra with distinguished scales, determined by parameters of
considered fields and their interaction.

In chaotic inflation scenario interaction of a Higgs field $\phi$
with inflaton $\eta$ can give rise to phase transitions on
inflationary stage, if this interaction induces positive mass term
$+\frac {\nu^2}{2} \eta^2 \phi^2$. When in the course of slow
rolling the amplitude of inflaton decreases below a certain critical
value $\eta_c = m/\nu$ the mass term in Higgs potential \beq
\label{Higgs} V(\phi, \eta)=-
\frac{m^2_{\phi}}{2}\phi^2+\frac{\lambda_{\phi}}{4}\phi^4 +\frac
{\nu^2}{2}\eta^2 \phi^2 \eeq changes sign and phase transition takes
place. Such phase transitions on inflationary stage lead to the
appearance of a characteristic spikes in the spectrum of initial
density perturbations. These spike--like perturbations, on scales
that cross the horizon ($60 \ge N \ge 1$) $e$-- folds before the end of
inflation reenter the horizon during the radiation or dust like era
and could in principle collapse to form primordial black holes. The
possibility of such spikes in chaotic inflation scenario was first
pointed out in Ref. \citen{KofLin} and realized  in Ref. \citen{Sakharov0} as a
mechanism of of PBH formation for the model of horizontal
unification\cite{Berezhiani1,Berezhiani2,Berezhiani3,Sakharov1}.

For vacuum expectation value of a Higgs field \beq\left\langle \phi
\right\rangle = \frac{m}{\lambda} = v\eeq and $\lambda \sim 10^{-3}$
the amplitude $\delta$ of spike in spectrum of density fluctuations,
generated in phase transition on inflationary stage is given by
\cite{Sakharov0} \beq \label{dspike}\delta \approx \frac{4}{9s}\eeq
with \beq \label{spike} s=\sqrt{ \frac{4}{9}+\kappa
10^5\left(\frac{v}{m_{pl}}\right)^2}-\frac{3}{2}, \eeq where $\kappa
\sim 1.$

If phase transition takes place at $e$--folding $N$ before the end
of inflation, the spike re-enters horizon on radiation dominance
(RD) stage and forms Black hole of mass \beq \label{Mrd} M \approx
\frac{m^2_{Pl}}{H_0} \exp\{2 N\}, \eeq where $H_0$ is the Hubble
constant in the period of inflation.

If the spike re-enters horizon on matter dominance (MD) stage it
should form black holes of mass \beq \label{Mmd} M \approx
\frac{m^2_{Pl}}{H_0} \exp\{3 N\}. \eeq

\subsection{First order
phase transitions as a source of black holes in the early
Universe}\label{phasetransitions} First order phase transition go
through bubble nucleation. Remind the common example of boiling
water. The simplest way to describe first order phase transitions
with bubble creation in early Universe is based on a scalar field
theory with two non degenerated vacuum states. Being stable at a
classical level, the false vacuum state decays due to quantum
effects, leading to a nucleation of bubbles of true vacuum and their
subsequent expansion \cite {6}. The potential energy of the false
vacuum is converted into a kinetic energy of bubble walls thus
making them highly relativistic in a short time. The bubble expands
till it collides with another one. As it was shown in Refs.
\citen{hawking3,5} a black hole may be created in a collision of
several bubbles. The probability for collision of two bubbles is
much higher. The opinion of the BH absence in such processes was
based on strict conservation of the original O(2,1) symmetry. As it
was shown in Refs. \citen{kkrs,kkrs1,kkrs2} there are ways to break it.
Firstly, radiation of scalar waves indicates the entropy increasing
and hence the permanent breaking of the symmetry during the bubble
collision. Secondly, the vacuum decay due to thermal fluctuation
does not possess this symmetry from the beginning. The
investigations \cite{kkrs,kkrs1,kkrs2} have shown that BH can be
created as well with a probability of order unity in collisions of
only two bubbles. It initiates an enormous production of BH that
leads to essential cosmological consequences discussed below.

Inflation models ended by a first order phase transition hold a
dignified position in the modern cosmology of early Universe (see
for example \cite{10,101,102,103,104,11,111}). The interest to these
models is due to, that such models are able to generate the observed
large-scale voids as remnants of the primordial bubbles for which
the characteristic wavelengths are several tens of Mpc\cite{11,111}.
A detailed analysis of a first order phase transition
in the context of extended inflation can be found in Ref.\citen{12}.
Hereafter we will be interested only in a final stage of inflation
when the phase transition is completed. Remind that a first order
phase transition is considered as completed immediately after
establishing of true vacuum percolation regime. Such regime is
established approximately when at least one bubble per unit Hubble
volume is nucleated. Accurate computation \cite{12} shows that first
order phase transition is successful if the following condition is
valid:
\begin{equation}
\label{14}Q\equiv \frac{4\pi }9\left( \frac \Gamma {H^4}\right)
_{t_{end}}=1.
\end{equation}
Here $\Gamma$ is the bubble nucleation rate. In the framework of
first order inflation models the filling of all space by true vacuum
takes place due to bubble collisions, nucleated at the final moment
of exponential expansion. The collisions between such bubbles occur
when they have comoving spatial dimension less or equal to the
effective Hubble horizon $H_{end}^{-1}$ at the transition epoch. If
we take $H_0=100hKm/\sec /Mpc$ in $\Omega =1$
Universe the comoving size of these bubbles is approximately $%
10^{-21}h^{-1}Mpc$. In the standard approach it believes that such
bubbles are rapidly thermalized without leaving a trace in the
distribution of matter and radiation. However, in the previous
subsection it has been shown that for any realistic parameters of
theory, the collision between only two bubble leads to BH creation
with the probability closely to 100\% . The mass of this BH is given
by\cite{kkrs,kkrs1,kkrs2}
\begin{equation}
\label{15}M_{BH}=\gamma _1M_{bub}
\end{equation}
where $\gamma _1\simeq 10^{-2}$ and $M_{bub}$ is the mass that could
be contained in the bubble volume at the epoch of collision in the
condition of a full thermalization of bubbles. The discovered
mechanism leads to a new direct possibility of PBH creation at the
epoch of reheating in first order inflation models. In standard
picture PBHs are formed in the early Universe if density
perturbations are sufficiently large, and the probability of PBHs
formation from small post- inflation initial perturbations is
suppressed (see subsection \ref{dust}). Completely different situation
takes place at final epoch of first order inflation stage; namely
collision between bubbles of Hubble size in percolation regime leads
to copious PBH formation with masses

\begin{equation}
\label{16}M_0=\gamma _1M_{end}^{hor}= \frac{\gamma
_1}2\frac{m_{pl}^2}{H_{end}},
\end{equation}
where $M_{end}^{hor}$ is the mass of Hubble horizon at the end of
inflation. According to (\ref{15}) the initial mass fraction of this
PBHs is given by $\beta _0\approx\gamma _1/e\approx 6\cdot 10^{-
3}$. For example, for typical value of $H_{end}\approx 4\cdot
10^{-6}m_{pl}$ the initial mass fraction $\beta $ is contained in
PBHs with mass $M_0\approx 1g$.

In general the Hawking evaporation of mini BHs could give rise to a
variety possible end states. It is generally assumed, that
evaporation proceeds until the PBH vanishes completely \cite{21},
but there are various arguments against this proposal (see e.g. Refs.
\citen{22,carr1,222,223}). If one supposes that BH evaporation leaves
a stable relic, then it is naturally to assume that it has a mass of
order $m_{rel}=km_{pl}$, where $ 1 \le k \le 10^2$. We can
investigate the consequences of PBH forming at the percolation epoch
after first order inflation, supposing that the stable relic is a
result of its evaporation. As it follows from the above
consideration the PBHs are preferentially formed with a typical mass
$M_0$ at a single time $t_1$. Hence the total density $\rho$ at this
time is
\begin{equation}
\label{totdens} \rho (t_1)=\rho_{\gamma}(t_1)+\rho_{PBH}(t_1)=
\frac{3(1-\beta_0)}{32\pi t_1^2}m_{pl}^2+ \frac{3\beta_0}{32\pi
t_1^2}m_{pl}^2,
\end{equation}
where $\beta_0$ denotes the fraction of the total density,
corresponding to PBHs in the period of their formation $t_1$. The
evaporation time scale can be written in the following form
\begin{equation}
\label{evop} \tau_{BH}=\frac{M_0^3}{g_*m_{pl}^4}
\end{equation}
where $g_*$ is the number of effective massless degrees of freedom.

Let us derive the density of PBH relics. There are two distinct
possibilities to consider.

The Universe is still radiation dominated (RD) at $\tau_{BH}$. This
situation will be hold if the following condition is valid
$\rho_{BH}(\tau_{BH})<\rho_{\gamma}(\tau_{BH})$. It is possible to
rewrite this condition in terms of Hubble constant at the end of
inflation
\begin{equation}
\label{con1} \frac{H_{end}}{m_{pl}}>\beta_0^{5/2}g_*^{-1/2}\simeq
10^{-6}
\end{equation}
Taking the present radiation density fraction of the Universe to be
$\Omega_{\gamma_0}=2.5\cdot 10^{-5}h^{-2}$ ($h$ being the Hubble
constant in the units of $100km\cdot s^{-1}Mpc^{-1}$), and using the
standard values for the present time and time when the density of
matter and radiation become equal, we find the contemporary
densities fraction of relics
\begin{equation}
\label{reldens} \Omega_{rel}\approx 10^{26}h^{-2}
k\left(\frac{H_{end}}{m_{pl}}\right)^{3/2}
\end{equation}
It is easily to see that relics overclose the Universe
($\Omega_{rel}>>1$) for any reasonable $k$ and
$H_{end}>10^{-6}m_{pl}$.

The second case takes place if the Universe becomes PBHs dominated
at period $t_1<t_2<\tau_{BH}$. This situation is realized under the
condition  $\rho_{BH}(t_2)<\rho_{\gamma}(t_2)$, which can be
rewritten in the form
\begin{equation}
\label{con2} \frac{H_{end}}{m_{pl}}<10^{-6}.
\end{equation}
The present day relics density fraction takes the form
\begin{equation}
\label{reldens2} \Omega_{rel}\approx 10^{28}h^{-2}
k\left(\frac{H_{end}}{m_{pl}}\right)^{3/2}
\end{equation}
Thus the Universe is not overclosed by relics only if the following
condition is valid
\begin{equation}
\label{con3} \frac{H_{end}}{m_{pl}}\le 2\cdot
10^{-19}h^{4/3}k^{-2/3}.
\end{equation}
This condition implies that the masses of PBHs created at the end of
inflation have to be larger then
\begin{equation}
\label{massr} M_0\ge 10^{11}g\cdot h^{-4/3}\cdot k^{2/3}.
\end{equation}
From the other hand there are a number of well--known cosmological
and astrophysical limits \cite{15,mujana,151,152,153,154,155} which
prohibit the creation of PBHs in the mass range (\ref{massr}) with
initial fraction of mass density close to $\beta_0\approx 10^{-2}$.

So one have to conclude that the effect of the false vacuum bag
mechanism of PBH formation makes impossible the coexistence of
stable remnants of PBH evaporation with the first order phase
transitions at the end of inflation.

\subsection{PBH evaporation as universal particle accelerator}\label{gravitino}

Presently there are no observational evidences, proving existence of
PBHs. However, even the absence of PBHs provides a very sensitive
theoretical tool to study physics of early Universe. PBHs represent
nonrelativistic form of matter and their density decreases with
scale factor $a$ as $\propto a^{-3} \propto T^{3}$, while the total
density is $\propto a^{-4}\propto T^{4}$ in the period of radiation
dominance (RD). Being formed within horizon, PBH of mass $M$, can be
formed not earlier than at
\begin{equation}\label{tfRD}t(M)=\frac{M}{m_{pl}}{t_{pl}}=\frac{M}{m_{pl}^2}.\end{equation} If they are formed
on RD stage, the smaller are the masses of PBHs, the larger becomes
their relative contribution to the total density on the modern MD
stage. Therefore, even the modest constraint for PBHs of mass $M$ on
their density \beq
\label{OmPBH}\Omega_{PBH}(M)=\frac{\rho_{PBH}(M)}{\rho_{c}}\eeq in
units of critical density $\rho_{c}=3 H^2/(8 \pi G)$ from the
condition that their contribution $\alpha(M)$ into the the total
density
\begin{equation}\label{defalpha}\alpha(M)\equiv\frac{\rho_{PBH}(M)}{\rho_{tot}}=\Omega_{PBH}(M)
\end{equation} for $\rho_{tot}=\rho_{c}$ does not exceed the density
of dark
matter\begin{equation}\label{DMalpha}\alpha(M)=\Omega_{PBH}(M) \le
\Omega_{DM}=0.23\end{equation} converts into a severe constraint on
this contribution
\begin{equation}\label{defbeta}\beta \equiv
\frac{\rho_{PBH}(M,t_f)}{\rho_{tot}(t_f)}\end{equation} in the
period $t_f$ of their formation. If formed on RD stage at
$t_f=t(M)$, given by (\ref{tfRD}), which corresponds to the
temperature $T_f=m_{pl}\sqrt{m_{pl}/M}$, PBHs contribute into the
total density in the end of RD stage at $t_{eq}$, corresponding to
$T_{eq}\approx 1 eV$,  by factor
$a(t_{eq})/a(t_f)=T_f/T_{eq}=m_{pl}/T_{eq}\sqrt{m_{pl}/M}$ larger,
than in the period of their formation. The constraint on $\beta(M)$,
following from Eq.(\ref{DMalpha}) is then given
by\begin{equation}\label{DMbeta}\beta(M)=\alpha(M)\frac{T_{eq}}{m_{pl}}\sqrt{\frac{M}{m_{pl}}}
\le 0.23 \frac{T_{eq}}{m_{pl}}\sqrt{\frac{M}{m_{pl}}}.\end{equation}

The possibility of PBH evaporation, revealed by S. Hawking
\cite{hawking4}, strongly influences effects of PBHs. In the strong
gravitational field near gravitational radius $r_g$ of PBH quantum
effect of creation of particles with momentum $p \sim 1/r_g$ is
possible. Due to this effect PBH turns to be a black body source of
particles with temperature (in the units
$\hbar=c=k=1$)\begin{equation}\label{TPBHev}T=\frac{1}{8\pi G
M}\approx10^{13} {\rm GeV} \frac{1 {\rm g}}{M}.\end{equation} The
evaporation timescale BH is $\tau_{BH} \sim M^3/m_{pl}^4$ (see
Eq.(\ref{evop}) and discussion in previous section) and at $M \le
10^{14}$~g is less, than the age of the Universe. Such PBHs can not
survive to the present time and the magnitude Eq.(\ref{DMalpha}) for
them should be re-defined and has the meaning of contribution to the
total density in the moment of PBH evaporation. For PBHs formed on
RD stage and evaporated on RD stage at $t<t_{eq}$ the relationship
Eq.(\ref{DMbeta}) between $\beta(M)$ and $\alpha(M)$ is given by
\cite{NovikovPBH,polnarev}
\begin{equation}\label{DMbetaRD}\beta(M)=\alpha(M)\frac{m_{pl}}{M}.\end{equation}
The relationship between $\beta(M)$ and $\alpha(M)$ has more
complicated form, if PBHs are formed on early dust-like stages
\cite{polnarev1,polnarev,khlopov6,book}, or such stages take place
after PBH formation\cite{khlopov6,book}. Relative contribution of
PBHs to total density does not grow on dust-like stage and the
relationship between $\beta(M)$ and $\alpha(M)$ depends on details
of a considered model. Minimal model independent factor
$\alpha(M)/\beta(M)$ follows from the account for enhancement,
taking place only during RD stage between the first second of
expansion and the end of RD stage at $t_{eq}$, since radiation
dominance in this period is supported by observations of light
element abundance and spectrum of CMB
\cite{polnarev1,polnarev,khlopov6,book}.

Effects of PBH evaporation make astrophysical data much more
sensitive to existence of PBHs.
 Constraining the abundance of primordial
black holes can lead to invaluable information on cosmological
processes, particularly as they are probably the only viable probe
for the power spectrum on very small scales which remain far from
the Cosmological Microwave Background (CMB) and Large Scale
Structures (LSS) sensitivity ranges. To date, only PBHs with initial
masses between $\sim 10^9$~g and $\sim 10^{16}$~g have led to
stringent limits (see {\it e.g.} Refs.
\citen{carr1,carrMG,LGreen,polnarev}) from consideration of the
entropy per baryon, the deuterium destruction, the $^4$He
destruction and the cosmic-rays currently emitted by the Hawking
process \cite{hawking4}. The existence of light PBHs should lead to
important observable constraints, either through the direct effects
of the evaporated particles (for initial masses between $10^{14}$~g
and $10^{16}$~g) or through the indirect effects of their
interaction with matter and radiation in the early Universe (for PBH
masses between $10^{9}$~g and $10^{14}$~g). In these constraints,
the effects taken into account are those related with known
particles. However, since the evaporation products are created by
the gravitational field, any quantum with a mass lower than the
black hole temperature should be emitted, independently of the
strength of its interaction. This could provide a copious production
of superweakly interacting particles that cannot not be in
equilibrium with the hot plasma of the very early Universe. It makes
evaporating PBHs a unique source of all the species, which can exist
in the Universe.

Following Refs. \citen{book,book3,khlopov6,khlopov7} and
\citen{lemoine,green1} (but in a different framework and using more
stringent constraints),
 limits on the mass fraction of black holes
at the time of their formation ($\beta \equiv
\rho_{PBH}/\rho_{tot}$) were derived in Ref. \citen{KBgrain} using the
production of gravitinos during the evaporation process. Depending
on whether gravitinos are expected to be stable or metastable, the
limits are obtained using the requirement that they do not overclose
the Universe and that the formation of light nuclei by the
interactions of $^4$He nuclei with nonequilibrium flux of D,T,$^3$He
and $^4$He does not contradict the observations. This approach is
more constraining than the usual study of photo-dissociation induced
by photons-photinos pairs emitted by decaying gravitinos. It opened
a new window for the upper limits on $\beta$ below $10^9$~g and correspondingly on
various mechanisms of PBH formation \cite{KBgrain}.

\subsection {Massive Primordial Black Holes - seeds for galaxy formation}\label{MBHwalls}

\subsubsection {Formation of closed walls in inflationary Universe}\label{walls}
To describe a mechanism for the appearance of massive walls of a
size essentially greater than the horizon at the end of inflation,
let us consider a complex scalar field with the
potential\cite{AGN,KRS,Khlopov:2004sc,book2}
\begin{equation}\label{V1} V(\varphi ) = \lambda (\left| \varphi
\right|^2  - f^2 /2)^2+\delta V(\theta ), \end{equation} where
$\varphi  = re^{i\theta } $. This field coexists with an inflaton
field which drives the Hubble constant $H$ during the inflational
stage. The term
\begin{equation} \label{L1} \delta V(\theta ) = \Lambda ^4 \left(
{1 - \cos \theta } \right), \end{equation} reflecting the
contribution of instanton effects to the Lagrangian renormalization
(see for example Ref. \citen{adams}), is negligible on the inflational
stage and during some period in the FRW expansion. The omitted term
(\ref{L1}) becomes significant, when temperature falls down the
values $T \sim \Lambda$. The mass of radial field component $r$ is
assumed to be sufficiently large with respect to $H$, which means
that the complex field is in the ground state even before the end of
inflation. Since the term (\ref{L1}) is negligible during inflation,
the field has the form $\varphi \approx f/\sqrt 2 \cdot e^{i\theta }
$, the quantity $f\theta$ acquiring the meaning of a massless field.


At the same time,  the well established behavior of quantum field
fluctuations on the de Sitter background \cite{Star80} implies that
the wavelength of a vacuum fluctuation of every scalar field grows
exponentially, having a fixed amplitude. Namely, when the wavelength
of a particular fluctuation, in the inflating Universe, becomes
greater than $H^{-1}$, the average amplitude of this fluctuation
freezes out at some  non-zero value because of the large friction
term in the equation of motion  of the scalar field, whereas its
wavelength grows exponentially. Such a frozen fluctuation is
equivalent to the appearance of a classical field that does not
vanish after averaging over macroscopic space intervals. Because the
vacuum must contain fluctuations of every wavelength, inflation
leads to the  creation of more and more new regions containing a
classical field of different amplitudes with scale greater than
$H^{-1}$. In the case of an effectively massless Nambu--Goldstone
field considered here, the averaged amplitude of phase fluctuations
generated during each e-fold (time interval $H^{-1}$)  is given by
\beq \label{fluctphase} \delta \theta = H/2\pi f. \eeq Let us assume
that the part of the Universe observed inside the contemporary
horizon $H_0^{-1}=3000h^{-1}$Mpc was inflating, over $N_U \simeq 60$
e-folds, out of a single causally connected domain of size $H^{-1}$,
which contains some average value of phase $\theta_0$ over it. When
inflation begins in this region, after one e-fold, the volume of the
Universe increases by a factor $e^3$ . The typical wavelength of the
fluctuation $\delta\theta$ generated during every e-fold is equal to
$H^{-1}$. Thus, the whole domain  $H^{-1}$, containing $\theta_{0}$,
after the first e-fold effectively becomes divided into  $e^3$
separate, causally disconnected domains of size $H^{-1}$. Each
domain contains almost homogeneous  phase value
$\theta_{0}\pm\delta\theta$. Thereby, more and more domains appear
with time, in which the phase differs significantly from the initial
value $\theta_0$. A principally important point is the appearance of
domains with phase $\theta >\pi$. Appearing only after a certain
period of time during which the Universe exhibited exponential
expansion, these domains turn out to be surrounded by a space with
phase $\theta <\pi$. The coexistence of domains with phases $\theta
<\pi$ and $\theta
>\pi$ leads, in the following, to formation of
a large-scale structure of topological defects.

The potential (\ref{V1}) possesses a $U(1)$ symmetry, which is
spontaneously broken, at least, after some period of inflation. Note
that the phase fluctuations during the first e-folds may, generally
speaking, transform eventually into fluctuations of the cosmic
microwave radiation, which will lead to imposing restrictions on the
scaling parameter $f$. This difficulty can be avoided by taking into
account the interaction of the field $\varphi$ with the inflaton
field (i.e. by making parameter $f$ a variable~\cite{book2}). This
spontaneous breakdown is holding by the condition of smallness of
the radial mass, $m_r=\sqrt{\lambda_{\phi}}>H$. At the same time the
condition \beq\label{angularmass} m_{\theta}=\frac{2f}{\Lambda}^2\ll
H \eeq on the angular mass provides the freezing out  of the phase
distribution until some moment of the FRW epoch.  After the
violation of condition (\ref{angularmass}) the term (\ref{L1})
contributes significantly to the potential (\ref{V1}) and explicitly
breaks the continuous symmetry along the angular direction. Thus,
potential (\ref{V1}) eventually has a number of discrete degenerate
minima in the angular direction at the points $\theta_{min}=0,\ \pm
2\pi ,\ \pm 4\pi,\ ...$ .

As soon as the angular mass $m_{\theta}$ is of the order of the
Hubble rate, the phase starts oscillating about the potential
minimum, initial values being different in various space domains.
Moreover, in the domains with the initial phase $\pi <\theta < 2\pi
$, the oscillations proceed around the potential minimum at $\theta
_{min}=2\pi$, whereas the phase in the surrounding space tends to a
minimum at the point $\theta _{min}=0$. Upon ceasing of the decaying
phase oscillations, the system contains domains characterized by the
phase $\theta _{min}=2\pi$ surrounded by space with $\theta
_{min}=0$. Apparently, on moving in any direction from inside to
outside of the domain, we will unavoidably pass through a point
where $\theta =\pi$ because the phase varies continuously. This
implies that a closed surface characterized by the phase $\theta
_{wall}=\pi$ must exist. The size of this surface depends on the
moment of domain formation in the inflation period, while the shape
of the surface may be arbitrary. The principal point for the
subsequent considerations is that the surface is closed. After
reheating of the Universe, the evolution of domains with the phase
$\theta >\pi $ proceeds on the background of the Friedman expansion
and is described by the relativistic equation of state. When the
temperature falls down to $T_* \sim \Lambda$, an equilibrium state
between the "vacuum" phase $\theta_{vac}=2\pi$ inside the domain and
the $\theta_{vac} =0$ phase outside it is established. Since the
equation of motion corresponding to potential (\ref{L1}) admits a
kink-like solution (see Ref. \citen{vs} and references therein), which
interpolates between two adjacent vacua $\theta_{vac} =0$  and
$\theta_{vac} =2\pi$,  a closed wall corresponding to the transition
region at $\theta =\pi$ is formed. The surface energy density of a
wall of width $\sim 1/m\sim  f/\Lambda^2$ is of the order of $\sim
f\Lambda ^2$ \footnote{The existence of such domain walls in theory
of the invisible axion was first pointed out in Ref.
\citen{sikivieinvisible}.}.

Note that if the coherent phase oscillations do not decay for a long
time, their energy density can play the role of CDM. This is the
case, for example, in the cosmology of the invisible axion (see
\cite{kim} and references therein).

It is clear that immediately after the end of inflation, the size of
domains which contains a phase $\theta_{vac} >2\pi$ essentially
exceeds the horizon size.  This situation is replicated in the size
distribution of vacuum walls, which appear at the temperature $T_*
\sim \Lambda$ whence the angular mass $m_{\theta}$ starts to build
up. Those walls, which are larger than the cosmological horizon,
still follow the general FRW expansion until the moment when they
get causally connected as a whole; this happens as soon as the size
of a wall becomes equal to the horizon size $R_h$. Evidently,
internal stresses developed in the wall after crossing  the horizon
initiate processes tending to minimize the wall  surface. This
implies that the wall tends, first, to acquire a  spherical shape
and, second, to contract toward the centre. For simplicity, we will
consider below the motion of closed spherical walls~\footnote{The
motion of closed vacuum walls has been driven analytically in
REfs. \citen{tkachev,sikivie}.}.

The wall energy is proportional to its area at the instant of
crossing the horizon. At the moment of maximum contraction, this
energy is almost completely converted into kinetic energy
\cite{Rubinwall}. Should the wall at the same moment be localized
within the gravitational radius, a PBH is formed.

Detailed consideration of BH formation was performed in Ref. \citen{AGN}.
The results of these calculations are sensitive to changes in the
parameter $\Lambda$ and the initial phase $\theta _U$. As the
$\Lambda$ value decreases to $\approx 1$GeV, still greater PBHs
appear with masses of up to $\sim 10^{40}$ g. A change in the
initial phase leads to sharp variations in the total number of black
holes.As was shown above, each domain generates a family of
subdomains in the close vicinity. The total mass of such a cluster
is only 1.5--2 times that of the largest initial black hole in this
space region. Thus, the calculations confirm the possibility of
formation of clusters of massive PBHs ( $\sim 100M_{\odot}$ and
above) in the earliest stages of the evolution of the Universe at a
temperature of $\sim 1-10$GeV. These clusters represent stable
energy density fluctuations around which increased baryonic (and
cold dark matter) density may concentrate in the subsequent stages,
followed by the evolution into galaxies.

It should be noted that additional energy density is supplied by
closed walls of small sizes. Indeed, because the smallness of their
gravitational radius, they do not collapse into BHs. After several
oscillations such walls disappear, leaving coherent fluctuations of
the PNG field. These fluctuations contribute to a local energy
density excess, thus facilitating the formation of galaxies.

The mass range of formed BHs is constrained by fundamental
parameters of the model $f$ and $\Lambda$. The maximal BH mass is
determined by the condition that the wall does not dominate locally
before it enters the cosmological horizon. Otherwise, local wall
dominance leads to a superluminal $a \propto t^2$ expansion for the
corresponding region, separating it from the other part of the
Universe. This condition corresponds to the mass \cite{book2}\beq
\label{Mmax} M_{max} =
\frac{m_{pl}}{f}m_{pl}(\frac{m_{pl}}{\Lambda})^2.\eeq The minimal
mass follows from the condition that the gravitational radius of BH
exceeds the width of wall and it is equal to\cite{KRS,book2}\beq
\label{Mmin} M_{min} = f(\frac{m_{pl}}{\Lambda})^2.\eeq

Closed wall collapse leads to primordial GW spectrum, peaked at \beq
\label{nupeak}\nu_0=3\cdot 10^{11}(\Lambda/f){\rm Hz} \eeq with
energy density up to \beq \label{OmGW}\Omega_{GW} \approx
10^{-4}(f/m_{pl}).\eeq At $f \sim 10^{14}$GeV this primordial
gravitational wave background can reach $\Omega_{GW}\approx
10^{-9}.$ For the physically reasonable values of \beq
1<\Lambda<10^8{\rm GeV}\eeq the maximum of spectrum corresponds to
\beq 3\cdot 10^{-3}<\nu_0<3\cdot 10^{5}{\rm Hz}.\eeq Another
profound signature of the considered scenario are gravitational wave
signals from merging of BHs in PBH cluster. These effects can
provide test of the considered approach in LISA experiment.
\section{Dark matter from flavor symmetry}
\label{flavor}
\subsection{Symmetry of known families}
The existence and observed properties of the three known quark-lepton families appeal to the broken $SU(3)_H$ family symmetry \cite{Berezhiani1,Berezhiani2,Berezhiani3}, which should be involved in the extension of the Standard model. It provides the possibility of the {\it Horizontal unification} in the "bottom-up" approach to the unified theory \cite{Sakharov1}. Even in its minimal implementation the model of {\it Horizontal unification} can reproduce the main necessary elements of the modern cosmology. It provides the physical mechanisms for inflation and baryosynthesis as well as it offers unified description of candidates for Cold, Warm, Hot and Unstable Dark Matter. Methods of cosmoparticle physics \cite{book,newBook} have provided the complete test of this model. Here we discuss the possibilities to link physical basis of modern cosmology to the parameters of broken family symmetry.

\subsubsection{Horizontal hierarchy}
The approach of Refs. \citen{Berezhiani1,Berezhiani2,Berezhiani3,Sakharov1} (and its revival in Refs. \citen{bai1,bai2,bai3}) follows the concept of local gauge symmetry $SU(3)_H$, first proposed by Chkareuli\cite{jon}. Under the action of this symmetry the left-handed quarks and leptons transform as $SU(3)_H$ triplets and the right-handed – as antitriplets. Their mass term transforms as
$3\bigotimes3=6\bigotimes \bar3$ and, therefore, can only form as a result of horizontal symmetry breaking.

This approach can be trivially extended to the case of $n$ generations, assuming the proper $SU(n)$ symmetry. For three generations, the choice of horizontal symmetry $SU(3)_H$ is the only possible choice because the orthogonal and vector-like gauge groups can not provide different representations for the left- and right-handed fermion states.

In the considered approach, the hypothesis that the structure of the mass matrix is determined by the structure of horizontal symmetry breaking, i.e., the structure of the vacuum expectation values of horizontal scalars carrying the $SU(3)_H$ breaking is justified.

The mass hierarchy between generations is related to the hypothesis of a hierarchy of such symmetry breaking. This hypothesis is called - the hypothesis of horizontal hierarchy (HHH)\cite{zurab1,zurab2,zurab3}.

The model is based on the gauge $SU(3)_H$ flavor symmetry, which is additional to the symmetry of the Standard model. It means that there exist 8 heavy horizontal gauge bosons and there are three multiplets of heavy Higgs fields $\xi_{ij}^{(n)}$ ($i$,$j$ - family indexes,$n=1,2,3$) in nontrivial (sextet or triplet) representations of $SU(3)_H$. These heavy Higgs bosons are singlets relative to electroweak symmetry and don't have Yukawa couplings with ordinary light fermions. They have direct coupling to heavy fermions. The latter are singlets relative to electroweak symmetry. Ordinary Higgs $\phi$ of the Standard model is singlet relative to $SU(3)_H$. It couples left-handed light fermions $f_L^i$ to their heavy right-handed partners $F_R^i$, which are coupled by heavy Higgses $\xi_{ij}$ with heavy left handed states $F_L^j$. Heavy left-handed states $F_L^j$ are coupled to right handed light states $f_R^j$ by a singlet scalar Higgs field $\eta$, which is singlet both relative to $SU(3)_H$ and electroweak group of symmetry. The described succession of transitions realizes Dirac see-saw mechanism, which reproduces the mass matrix $m_{ij}$ of ordinary light quarks and charged leptons $f$ due to mixing with their heavy partners $F$. It fixes the ratio of vacuum expectation values of heavy Higgs fields, leaving their absolute value as the only main free parameter, which is determined from analysis of physical, astrophysical and cosmological consequences.

The $SU(3)_H$ flavor symmetry should be chiral to eliminate the flavor symmetric mass term. The condition of absence of anomalies implies heavy partners of light neutrinos, and the latter acquire mass by Majorana see-saw mechanism. The natural absence in the heavy Higgs potentials of triple couplings, which do not appear as radiative effects of any other (gauge or Yukawa) interaction, supports additional global U(1) symmetry, which can be associated with Peccei-Quinn symmetry and whose breaking results in the Nambu-Goldstone scalar filed, which shares the properties of axion, Majoron and singlet familon.
\subsubsection{Horizontal unification}
The model provides complete test (in which its simplest implementation is already ruled out) in a combination of laboratory tests and analysis of cosmological and astrophysical effects. The latter include the study of the effect of radiation of axions on the processes of stellar evolution, the study of the impact of the effects of primordial axion fields and massive unstable neutrino on the dynamics of formation of the large-scale structure of the Universe, as well as analysis of the mechanisms of inflation and baryosynthesis based on the physics of the hidden sector of the model.

The model results in physically self-consistent inflationary scenarios with dark matter in the baryon-asymmetric Universe. In these scenarios, all steps of the cosmological evolution correspond quantitatively to the parameters of particle theory.
The physics of the inflaton corresponds to the Dirac 'see-saw' mechanism of generation of the mass of the quarks and charged leptons, leptogenesis  of baryon asymmetry is based on the physics of Majorana neutrino masses. The parameters of axion CDM, as well as the masses and lifetimes of neutrinos correspond to the hierarchy of breaking of the $SU(3)_H$ symmetry of families.
\subsection{Stable charged constituents of Dark Atoms}\label{asymmetry}
New stable particles may possess new U(1)
gauge charges and bind by Coulomb-like forces in composite dark
matter species. Such dark atoms would look nonluminous, since they
radiate invisible light of U(1) photons. Historically mirror matter
(see subsubsection \ref{mirror} and Refs. \citen{book,OkunRev} for review and references) seems to be the
first example of such a nonluminous atomic dark matter.

However, it turned out that the possibility of new stable charged leptons and quarks is not completely excluded and Glashow's tera-helium\cite{Glashow} has offered a new solution for
dark atoms of dark matter. Tera-U-quarks with electric charge +2/3
formed stable (UUU) +2 charged "clusters" that formed with two -1
charged tera-electrons E neutral [(UUU)EE] tera-helium "atoms" that
behaved like Weakly Interacting Massive Particles (WIMPs). The main
problem for this solution was to suppress the abundance of
positively charged species bound with ordinary electrons, which
behave as anomalous isotopes of hydrogen or helium. This problem
turned to be unresolvable\cite{BKSR1}, since the model\cite{Glashow}
predicted stable tera-electrons $E^-$ with charge -1.
As soon as primordial helium is formed in the Standard Big Bang
Nucleosynthesis (SBBN) it captures all the free $E^-$ in positively
charged $(He E)^+$ ion, preventing any further suppression of
positively charged species. Therefore, in order to avoid anomalous
isotopes overproduction, stable particles with charge -1 (and
corresponding antiparticles) should be absent, so that stable
negatively charged particles should have charge -2 only.

Elementary particle frames for heavy stable -2 charged species are
provided by: (a) stable "antibaryons" $\bar U \bar U \bar U$ formed
by anti-$U$ quark of fourth generation\cite{Q,I,BKSR4,Belotsky:2008se,DADM}
(b) AC-leptons\cite{DADM,FKS}, predicted in the
extension \cite{FKS} of standard model, based on the approach of
almost-commutative geometry\cite{bookAC}.  (c) Technileptons and
anti-technibaryons \cite{KK} in the framework of walking technicolor
models (WTC)\cite{Sannino:2004qp,Hong:2004td,Dietrich:2005jn,Dietrich:2005wk,Gudnason:2006ug,Gudnason:2006yj}. (d) Finally, stable charged
clusters $\bar u_5 \bar u_5 \bar u_5$ of (anti)quarks $\bar u_5$ of
5th family can follow from the approach, unifying spins and charges\cite{Norma,Norma2,Norma3,Norma4,Norma5}. Since all these models also predict corresponding +2
charge antiparticles, cosmological scenario should provide mechanism
of their suppression, what can naturally take place in the
asymmetric case, corresponding to excess of -2 charge species,
$O^{--}$. Then their positively charged antiparticles can
effectively annihilate in the early Universe.

If new stable species belong to non-trivial representations of
electroweak SU(2) group, sphaleron transitions at high temperatures
can provide the relationship between baryon asymmetry and excess of
-2 charge stable species, as it was demonstrated in the case of WTC
in Refs. \citen{KK,Levels1,KK2,unesco,iwara,I2}.


\subsubsection{Problem of tera-fermion composite dark matter}
Glashow's Tera-helium Universe was first inspiring example of the
composite dark matter scenario. $SU(3)_c \times SU(2) \times SU(2)'
\times U(1)$ gauge model\cite{Glashow} was aimed to explain the origin of the neutrino mass and to solve the problem of strong CP-violation in QCD. New extra $SU(2)'$ symmetry acts on three heavy generations of
tera-fermions  linked with the light fermions by $CP'$
transformation. $SU(2)'$ symmetry breaking at TeV scale makes
tera-fermions much heavier than their light partners. Tera-fermion
mass spectrum is the same as for light generations, but all the
masses are scaled by the same factor of about $10^6$. Thus the
masses of lightest heavy particles are in {\it tera}-eV (TeV) range,
explaining their name.

Glashow's model\cite{Glashow} takes into account
that
 very heavy quarks $Q$ (or antiquarks $\bar Q$) can form bound states with other heavy quarks
 (or antiquarks) due to their Coulomb-like QCD attraction, and the binding energy of these states
 substantially exceeds the binding energy of QCD confinement.
Then stable $(QQq)$ and $(QQQ)$ baryons can exist.

According to Ref. \citen{Glashow} primordial heavy quark $U$ and heavy
electron $E$ are stable and
may form a neutral $(UUUEE)$ "atom"
with $(UUU)$ hadron as nucleus and two $E^-$s as "electrons". The
gas of such "tera-helium atoms" was proposed in Ref. \citen{Glashow} as a candidate for a
WIMP-like dark matter.

The problem of such scenario is an
inevitable presence of "products of incomplete combustion" and the
necessity to decrease their abundance.

Unfortunately, as it was shown in Ref. \citen{BKSR1}, this
picture of Tera-helium Universe can not be realized.

When ordinary $^4$He is formed in Big Bang
Nucleosynthesis, it binds all the free
$E^-$ into positively charged $(^4HeE^-)^+$ "ions". This puts
Coulomb barrier for any successive $E^-E^+$ annihilation or any
effective $EU$ binding. It removes  a possibility to suppress the abundance of
unwanted tera-particle species (like $(eE^+)$, $(^4He Ee)$ etc).
For instance the remaining abundance of $(eE^+)$ and $(^4HeE^-e)$ exceeds the terrestrial upper limit for anomalous hydrogen by
{\it 27 orders} of magnitude\cite{BKSR1}.

\subsubsection{Composite dark matter from almost commutative geometry}
The AC-model is based on the specific mathematical approach of
unifying general relativity, quantum mechanics and gauge symmetry\cite{FKS,bookAC}.
This realization naturally embeds the Standard model, both
reproducing its gauge symmetry and Higgs mechanism with prediction of a Higgs boson mass. AC model
 is in some sense alternative to SUSY, GUT and superstring extension of Standard model. The AC-model\cite{FKS} extends the fermion content of the Standard
model by two heavy particles, $SU(2)$ electro-weak singlets, with opposite electromagnetic charges.
Each of them has its own antiparticle. Having no other gauge charges of Standard model,
these particles (AC-fermions) behave as heavy stable leptons with
charges $-2e$ and $+2e$, called $A^{--}$ and $C^{++}$, respectively.

Similar to the Tera-helium Universe, AC-lepton relics from
intermediate stages of a multi-step process towards a final $(AC)$
atom formation must survive in the present Universe. In spite of the assumed excess of
particles ($A^{--}$ and $C^{++}$) the abundance of relic
antiparticles ($\bar A^{++}$ and $\bar C^{--}$) is not negligible.
There may be also a significant fraction of $A^{--}$ and $C^{++}$, which remains
unbound after recombination process of these particles into $(AC)$ atoms took place. As soon as $^4He$ is formed in Big
Bang nucleosynthesis, the primordial component of free anion-like AC-leptons
($A^{--}$) is mostly trapped in the first three minutes into a
neutral O-helium atom $^4He^{++}A^{--}$.
O-helium is able to capture free $C^{++}$ creating $(AC)$ atoms and releasing $^4He$ back. In the same way the annihilation of antiparticles speeds up. $C^{++}$-O-helium reactions stop, when their timescale exceeds a cosmological time, leaving O-helium and $C^{++}$ relics in the Universe. The catalytic reaction of O-helium with $C^{++}$ in the dense matter bodies provides successive
$(AC)$ binding that suppresses terrestrial
anomalous isotope abundance below the experimental upper limit. Due to screened charge of AC-atoms they have WIMP-like interaction with the ordinary matter. Such WIMPs are inevitably accompanied by a tiny component of nuclear interacting O-helium.

\subsubsection{Stable charged techniparticles in Walking Technicolor}

The minimal walking technicolor model\cite{Sannino:2004qp,Hong:2004td,Dietrich:2005jn,Dietrich:2005wk,Gudnason:2006ug,Gudnason:2006yj}
has two techniquarks, i.e. up $U$ and down $D$, that transform
under the adjoint representation of an $SU(2)$ technicolor gauge
group. The six
Goldstone bosons $UU$, $UD$, $DD$ and their corresponding
antiparticles carry technibaryon number since they are made of
two techniquarks or two anti-techniquarks. This means that if there is no
processes violating the technibaryon number the lightest
technibaryon will be stable.

The electric charges of $UU$, $UD$,
and $DD$ are given in general by $q+1$, $q$, and $q-1$
respectively, where $q$ is an arbitrary real number. The model requires in addition
the existence of a fourth family of leptons, i.e. a ``new
neutrino'' $\nu'$ and a ``new electron'' $\zeta$. Their electric charges are in
terms of $q$ respectively $(1-3q)/2$ and $(-1-3q)/2$.

There are three possibilities for a scenario of dark atoms of dark matter. The first one is to have an excess of $\bar{U}\bar{U}$ (charge $-2$).
The technibaryon
number $TB$ is conserved and therefore $UU$ (or $\bar{U}\bar{U}$) is
stable. The second possibility is to
have excess of $\zeta$ that also has $-2$ charge and is
stable, if $\zeta$ is lighter than $\nu'$ and technilepton number $L'$  is conserved. In the both cases
stable particles with $-2$ electric charge have substantial relic
densities and can capture $^4He^{++}$ nuclei to form a neutral techni-O-helium
atom.
Finally there is a
possibility to have both  $L'$ and $TB$ conserved. In this case, the dark matter would be composed
of bound atoms $(^4He^{++}\zeta^{--})$ and $(\zeta^{--}(U U )^{++})$. In the latter case the excess of $\zeta^{--}$ should be larger, than the excess of $(U U )^{++})$, so that WIMP-like $(\zeta^{--}(U U )^{++})$ is subdominant at the dominance of nuclear interacting techni-O-helium.

The technicolor and the
Standard Model particles are in thermal equilibrium as long as the
timescale of the weak (and color) interactions is smaller than the
cosmological time. The sphalerons allow violation of  $TB$, of baryon number $B$, of lepton number $L$ and  $L'$ as
long as the temperature of the Universe exceeds the electroweak scale.
It was shown in\cite{KK} that there is a balance between the excess of techni(anti)baryons, $(\bar{U}\bar{U})^{--}$,
technileptons $\zeta^{--}$ or of the both over the corresponding particles ($UU$ and/or $\zeta^{++}$) and the observed baryon asymmetry
of the Universe. It was also shown the there are parameters of the model, at which this asymmetry has
proper sign and value, explaining the dark matter density.

\subsubsection{\label{4generation} Stable particles of 4th generation matter}
Modern precision data
on the parameters of the Standard model do not exclude\cite{Maltoni:1999ta}
the existence of
the  4th generation of quarks and leptons. The 4th generation follows from heterotic string phenomenology and
its difference from the three known light generations can be
explained by a new conserved charge, possessed only by
its quarks and leptons\cite{Q,I,Belotsky:2000ra,Belotsky:2005uj,Belotsky:2004st}. Strict conservation of this charge makes the
lightest particle of 4th family (neutrino) absolutely
stable, but it was shown in Refs. \citen{Belotsky:2000ra,Belotsky:2005uj,Belotsky:2004st} that this neutrino cannot be the dominant form of the dark matter.
The same conservation law requires the lightest quark to be long living
\cite{Q,I}. In principle the lifetime of $U$ can exceed the age of the
Universe, if $m_U<m_D$\cite{Q,I}.
Provided that sphaleron transitions establish excess of $\bar U$ antiquarks at the observed baryon asymmetry
 $(\bar U \bar U \bar U)$ can be formed and bound with $^4He$ in atom-like state
of O-helium\cite{I}.

In the successive discussion of OHe dark matter we generally don't specify the type of $-2$ charged particle, denoting it as $O^{--}$.
However, one should note that the AC model doesn't provide OHe as the dominant form of dark matter, so that the quantitative features of OHe dominated Universe are not related to this case.
\section{Dark atoms with helium shell}
\label{ohe}
Here we concentrate on the properties of OHe atoms, their interaction with matter and qualitative picture of OHe cosmological evolution\cite{I,Levels,FKS,KK,unesco,Khlopov:2008rp,KhlopovPHE} and observable effects. We show following Refs. \citen{DADM,DMDA} that interaction of OHe with nuclei in
underground detectors can  explain positive results
of dark matter searches in DAMA/NaI (see for review Ref. \citen{DAMA-review})
and DAMA/LIBRA\cite{Bernabei:2008yi}
experiments by annual modulations of radiative capture of O-helium, resolving the controversy
between these results and the results of other experimental groups.

After it is formed
in the Standard Big Bang Nucleosynthesis (SBBN), $^4He$ screens the excessive
$O^{--}$ charged particles in composite $(^4He^{++}O^{--})$ {\it
O-helium} ($OHe$) ``atoms''\cite{I}.

In all the considered forms of O-helium, $O^{--}$ behaves either as lepton or
as specific "heavy quark cluster" with strongly suppressed hadronic
interaction. Therefore O-helium interaction with matter is
determined by nuclear interaction of $He$. These neutral primordial
nuclear interacting species can play the role of a nontrivial form of strongly
interacting dark matter\cite{Starkman,Wolfram,Starkman2,Javorsek,Mitra,Mack,McGuire:2001qj,McGuire2,ZF}, giving rise to a Warmer than
Cold dark matter scenario\cite{Levels,Levels1,KK2}.
\subsection{OHe atoms and their interaction with nuclei}
The structure of OHe atom follows from the general
analysis of the bound states of $O^{--}$ with nuclei.

Consider a simple model\cite{Cahn,Pospelov,Kohri}, in which the nucleus is
regarded as a sphere with uniform charge density and in which the
mass of the $O^{--}$ is assumed to be much larger than that of the
nucleus. Spin dependence is also not taken into account so that both
the particle and nucleus are considered as scalars. Then the
Hamiltonian is given by
\begin{equation}
    H=\frac{p^2}{2 A m_p} - \frac{Z Z_x \alpha}{2 R} + \frac{Z Z_x \alpha}{2 R} \cdot (\frac{r}{R})^2,
\end{equation}
for short distances $r<R$ and
\begin{equation}
    H=\frac{p^2}{2 A m_p} - \frac{Z Z_x \alpha}{R},
\end{equation}
for long distances $r>R$, where $\alpha$ is the fine structure
constant, $R = d_o A^{1/3} \sim 1.2 A^{1/3} /(200 MeV)$ is the
nuclear radius, $Z$ is the electric charge of nucleus and $Z_x=2$ is
the electric charge of negatively charged particle $X^{--}$. Since
$A m_p \ll M_X$ the reduced mass is $1/m= 1/(A m_p) + 1/M_X \approx
1/(A m_p)$.

For small nuclei the Coulomb binding energy is like in hydrogen atom
and is given by
\begin{equation}
    E_b=\frac{1}{2} Z^2 Z_x^2 \alpha^2 A m_p.
\end{equation}

For large nuclei $X^{--}$ is inside nuclear radius and the harmonic
oscillator approximation is valid for the estimation of the binding
energy
\begin{equation}
    E_b=\frac{3}{2}(\frac{Z Z_x \alpha}{R}-\frac{1}{R}(\frac{Z Z_x \alpha}{A m_p R})^{1/2}).
\label{potosc}
\end{equation}

For the intermediate regions between these two cases with the use of
trial function of the form $\psi \sim e^{- \gamma r/R}$ variational
treatment of the problem\cite{Cahn,Pospelov,Kohri} gives
\begin{equation}
    E_b=\frac{1}{A m_p R^2} F(Z Z_x \alpha A m_p R ),
\end{equation}
where the function $F(a)$ has limits
\begin{equation}
    F(a \rightarrow 0) \rightarrow \frac{1}{2}a^2  - \frac{2}{5} a^4
\end{equation}
and
\begin{equation}
    F(a \rightarrow \infty) \rightarrow \frac{3}{2}a  - (3a)^{1/2},
\end{equation}
where $a = Z Z_x \alpha A m_p R$. For $0 < a < 1$ the Coulomb model
gives a good approximation, while at $2 < a < \infty$ the harmonic
oscillator approximation is appropriate.

In the case of OHe $a = Z Z_x \alpha A m_p R \le 1$, what proves its
Bohr-atom-like structure, assumed in Refs. \citen{I,KK,unesco,iwara,I2}.
The radius of Bohr orbit in these ``atoms"
\cite{I,Levels} $r_{o} \sim 1/(Z_{o} Z_{He}\alpha m_{He}) \approx 2
\cdot 10^{-13} \cm $.
However, the size of
He nucleus, rotating around $O^{--}$ in this Bohr atom, turns out to be of
the order and even a bit larger than the radius $r_o$ of its Bohr
orbit, and the corresponding correction to the binding energy due to
non-point-like charge distribution in He is significant.

Bohr atom like structure of OHe seems to provide a possibility to
use the results of atomic physics for description of OHe interaction
with matter. However, the situation is much more complicated. OHe
atom is similar to the hydrogen, in which electron is hundreds times
heavier, than proton, so that it is proton shell that surrounds
"electron nucleus". Nuclei that interact with such "hydrogen" would
interact first with strongly interacting "protonic" shell and such
interaction can hardly be treated in the framework of perturbation
theory. Moreover in the description of OHe interaction the account
for the finite size of He, which is even larger than the radius of
Bohr orbit, is important. One should consider, therefore, the
analysis, presented below, as only a first step approaching true
nuclear physics of OHe.

The approach of Refs. \citen{Levels,Levels1} assumes the following
picture of OHe interaction with nuclei: OHe is a neutral atom in the ground state,
perturbed  by Coulomb and nuclear forces of the approaching nucleus.
The sign of OHe polarization changes with the distance: at larger distances Stark-like effect takes place - nuclear Coulomb force polarizes OHe so that  nucleus is attracted by the induced dipole moment of OHe, while as soon as the perturbation by nuclear force starts to dominate the nucleus polarizes OHe in the opposite way so that He is situated more close to the nucleus, resulting in the repulsive effect of the helium shell of OHe.
When helium is completely merged with the nucleus the interaction is
reduced to the oscillatory potential of $O^{--}$ with
homogeneously charged merged nucleus with the charge $Z+2$.

\begin{figure}
    \begin{center}
        \includegraphics[scale=0.4]{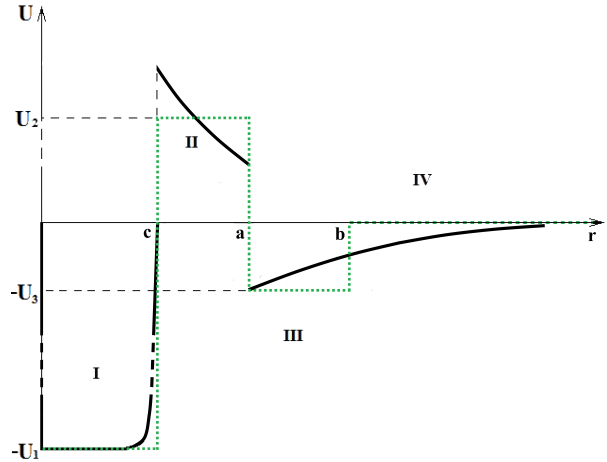}
        \caption{The potential of OHe-nucleus system and its rectangular well approximation.}
        \label{pic1}
    \end{center}
\end{figure}

\begin{figure}
\begin{center}
\includegraphics[scale=0.7]{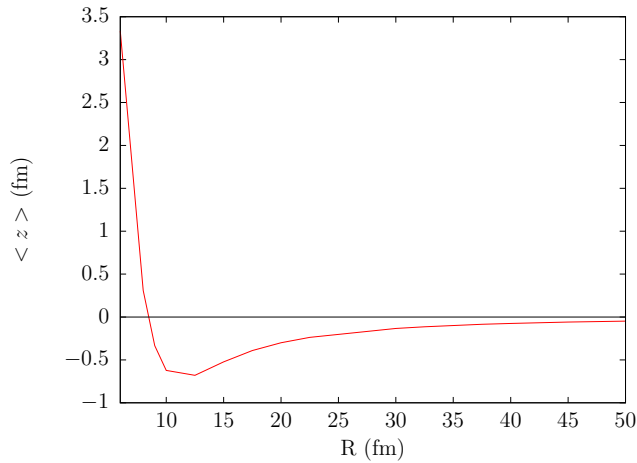}
\caption{Polarization $<z>$ (Fm) of OHe as a function of the distance
$R$ (fm) of an external sodium nucleus, calculated in Ref. \citen{quentin} in the framework of perturbation theory.}
\label{Pol}
 \end{center}
\end{figure}

Therefore OHe-nucleus potential can have qualitative feature, presented on Fig.~\ref{pic1}:
the potential well at large distances (regions III-IV) is changed by a potential wall in region II. The existence of this potential barrier is crucial for all the qualitative features of OHe scenario: it causes suppression of reactions with transition of OHe-nucleus system to levels in the potential well of the region I, provides the dominance of elastic scattering while transitions to levels in the shallow well (regions III-IV) should dominate in reactions of OHe-nucleus capture. The proof of this picture implies accurate and detailed quantum-mechanical treatment, which was started in Ref. \citen{quentin}. With the use of perturbation theory it was shown that OHe polarization changes sign, as the nucleus approaches OHe (as it is given on Fig. \ref{Pol}), but the perturbation approach was not valid for the description at smaller distances, while the estimations indicated that this change of polarization may not be sufficient for creation of the potential, given by Fig.~\ref{pic1}. If the picture of Fig.~\ref{pic1} is not proved, one may need more sophisticated models retaining the ideas of OHe scenario, which involve more elements of new physics, as proposed in Ref. \citen{Wallemacq:2013hsa}.

On the other hand, O-helium, being an $\alpha$-particle with screened electric charge,
can catalyze nuclear transformations, which can influence primordial
light element abundance and cause primordial heavy element
formation. It is especially important for quantitative estimation of
role of OHe in Big Bang Nucleosynthesis and in stellar evolution.
These effects need a special detailed and complicated
study of OHe nuclear physics and
this work is under way.

The qualitative
picture of OHe cosmological evolution is presented below following Refs. \citen{I,Levels,FKS,KK,Levels1,unesco,Khlopov:2008rp,DADM}
and is based on the idea of the dominant role of elastic
collisions in OHe interaction with baryonic matter.

\subsection{Large Scale structure formation by OHe dark matter}
Due to elastic nuclear interactions of its helium constituent with nuclei in
the cosmic plasma, the O-helium gas is in thermal equilibrium with
plasma and radiation on the Radiation Dominance (RD) stage, while
the energy and momentum transfer from plasma is effective. The
radiation pressure acting on the plasma is then transferred to
density fluctuations of the O-helium gas and transforms them in
acoustic waves at scales up to the size of the horizon.

At temperature $T < T_{od} \approx 1 S_3^{2/3}\eV$ the energy and
momentum transfer from baryons to O-helium is not effective
\cite{I,KK} because $$n_B \sv (m_p/m_o) t < 1,$$ where $m_o$ is the
mass of the $OHe$ atom and $S_3= m_o/(1 \TeV)$. Here \beq \sigma
\approx \sigma_{o} \sim \pi r_{o}^2 \approx
10^{-25}\cm^2\label{sigOHe}, \eeq and $v = \sqrt{2T/m_p}$ is the
baryon thermal velocity. Then O-helium gas decouples from plasma. It
starts to dominate in the Universe after $t \sim 10^{12}\s$  at $T
\le T_{RM} \approx 1 \eV$ and O-helium ``atoms" play the main
dynamical role in the development of gravitational instability,
triggering the large scale structure formation. The composite nature
of O-helium determines the specifics of the corresponding dark
matter scenario.

At $T > T_{RM}$ the total mass of the $OHe$ gas with density $\rho_d
= (T_{RM}/T) \rho_{tot} $ is equal to
$$M=\frac{4 \pi}{3} \rho_d t^3 = \frac{4 \pi}{3} \frac{T_{RM}}{T} m_{Pl}
(\frac{m_{Pl}}{T})^2$$ within the cosmological horizon $l_h=t$. In
the period of decoupling $T = T_{od}$, this mass  depends strongly
on the O-helium mass $S_3$ and is given by \cite{KK}\beq M_{od} =
\frac{T_{RM}}{T_{od}} m_{Pl} (\frac{m_{Pl}}{T_{od}})^2 \approx 2
\cdot 10^{44} S^{-2}_3 \g = 10^{11} S^{-2}_3 M_{\odot}, \label{MEPm}
\eeq where $M_{\odot}$ is the solar mass. O-helium is formed only at
$T_{o}$ and its total mass within the cosmological horizon in the
period of its creation is $M_{o}=M_{od}(T_{od}/T_{o})^3 = 10^{37}
\g$.

On the RD stage before decoupling, the Jeans length $\lambda_J$ of
the $OHe$ gas was restricted from below by the propagation of sound
waves in plasma with a relativistic equation of state
$p=\epsilon/3$, being of the order of the cosmological horizon and
equal to $\lambda_J = l_h/\sqrt{3} = t/\sqrt{3}.$ After decoupling
at $T = T_{od}$, it falls down to $\lambda_J \sim v_o t,$ where $v_o
= \sqrt{2T_{od}/m_o}.$ Though after decoupling the Jeans mass in the
$OHe$ gas correspondingly falls down
$$M_J \sim v_o^3 M_{od}\sim 3 \cdot 10^{-14}M_{od},$$ one should
expect a strong suppression of fluctuations on scales $M<M_o$, as
well as adiabatic damping of sound waves in the RD plasma for scales
$M_o<M<M_{od}$. It can provide some suppression of small scale
structure in the considered model for all reasonable masses of
O-helium. The significance of this suppression and its effect on the
structure formation needs a special study in detailed numerical
simulations. In any case, it can not be as strong as the free
streaming suppression in ordinary Warm Dark Matter (WDM) scenarios,
but one can expect that qualitatively we deal with Warmer Than Cold
Dark Matter model.

At temperature $T < T_{od} \approx 1 S_3^{2/3} \keV$ the energy and
momentum transfer from baryons to O-helium is not effective\cite{I,Levels,Levels1}
and O-helium gas decouples from plasma. It
starts to dominate in the Universe after $t \sim 10^{12}\s$  at $T
\le T_{RM} \approx 1 \eV$ and O-helium ``atoms" play the main
dynamical role in the development of gravitational instability,
triggering the large scale structure formation. The composite nature
of O-helium determines the specifics of the corresponding warmer than cold dark
matter scenario.

Being decoupled from baryonic matter, the $OHe$ gas does not follow
the formation of baryonic astrophysical objects (stars, planets,
molecular clouds...) and forms dark matter halos of galaxies. It can
be easily seen that O-helium gas is collisionless for its number
density, saturating galactic dark matter. Taking the average density
of baryonic matter one can also find that the Galaxy as a whole is
transparent for O-helium in spite of its nuclear interaction. Only
individual baryonic objects like stars and planets are opaque for
it.

\subsection{Anomalous component of cosmic rays}
O-helium atoms can be destroyed in astrophysical processes, giving
rise to acceleration of free $O^{--}$ in the Galaxy.

O-helium can be ionized due to nuclear interaction with cosmic rays\cite{I,I2}.
Estimations\cite{I,Mayorov} show that for the number
density of cosmic rays $ n_{CR}=10^{-9}\cm^{-3}$ during the age of
Galaxy a fraction of about $10^{-6}$ of total amount of OHe is
disrupted irreversibly, since the inverse effect of recombination of
free $O^{--}$ is negligible. Near the Solar system it leads to
concentration of free $O^{--}$ $ n_{O}= 3 \cdot 10^{-10}S_3^{-1}
\cm^{-3}.$ After OHe destruction free $O^{--}$ have momentum of
order $p_{O} \cong \sqrt{2 \cdot m_{o} \cdot I_{o}} \cong 2 \GeV
S_3^{1/2}$ and velocity $v/c \cong 2 \cdot 10^{-3} S_3^{-1/2}$ and
due to effect of Solar modulation these particles initially can
hardly reach Earth\cite{KK2,Mayorov}. Their acceleration by Fermi
mechanism or by the collective acceleration forms power spectrum of
$O^{--}$ component at the level of $O/p \sim n_{O}/n_g = 3 \cdot
10^{-10}S_3^{-1},$ where $n_g \sim 1 \cm^{-3}$ is the density of
baryonic matter gas.

At the stage of red supergiant stars have the size $\sim 10^{15}
\cm$ and during the period of this stage$\sim 3 \cdot 10^{15} \s$,
up to $\sim 10^{-9}S_3^{-1}$ of O-helium atoms per nucleon can be
captured\cite{KK2,Mayorov}. In the Supernova explosion these OHe
atoms are disrupted in collisions with particles in the front of
shock wave and acceleration of free $O^{--}$ by regular mechanism
gives the corresponding fraction in cosmic rays. However, this
picture needs detailed analysis, based on the development of OHe
nuclear physics and numerical studies of OHe evolution in the
stellar matter.

If these mechanisms of $O^{--}$ acceleration are effective, the
anomalous low $Z/A$ component of $-2$ charged $O^{--}$ can be
present in cosmic rays at the level $O/p \sim n_{O}/n_g \sim
10^{-9}S_3^{-1},$ and be within the reach for PAMELA and AMS02
cosmic ray experiments.

In the framework of Walking Technicolor model the excess of both
stable $\zeta^{--}$ and $(UU)^{++}$ is possible\cite{KK2}, the latter
being two-three orders of magnitude smaller, than the former. It
leads to the two-component composite dark matter scenario with the
dominant OHe accompanied by a subdominant WIMP-like component of
$(\zeta^{--}(U U )^{++})$ bound systems. Technibaryons can
be metastable and decays of $(UU)^{++}$ can provide
explanation for anomalies, observed in high energy cosmic positron
spectrum by PAMELA, FERMI-LAT and AMS02.

\subsection{Positron annihilation and gamma lines in galactic
bulge}
Inelastic interaction of O-helium with the matter in the
interstellar space and its de-excitation can give rise to radiation
in the range from few keV to few  MeV. In the galactic bulge with
radius $r_b \sim 1 \kpc$ the number density of O-helium can reach
the value $n_o\approx 3 \cdot 10^{-3}/S_3 \cm^{-3}$ and the
collision rate of O-helium in this central region was estimated in
\cite{I2}: $dN/dt=n_o^2 \sigma v_h 4 \pi r_b^3 /3 \approx 3 \cdot
10^{42}S_3^{-2} \s^{-1}$. At the velocity of $v_h \sim 3 \cdot 10^7
\cm/\s$ energy transfer in such collisions is $\Delta E \sim 1 \MeV
S_3$. These collisions can lead to excitation of O-helium. If 2S
level is excited, pair production dominates over two-photon channel
in the de-excitation by $E0$ transition and positron production with
the rate $3 \cdot 10^{42}S_3^{-2} \s^{-1}$ is not accompanied by
strong gamma signal. According to Ref. \citen{Finkbeiner:2007kk} this rate
of positron production for $S_3 \sim 1$ is sufficient to explain the
excess in positron annihilation line from bulge, measured by
INTEGRAL (see Ref. \citen{integral} for review and references). If $OHe$
levels with nonzero orbital momentum are excited, gamma lines should
be observed from transitions ($ n>m$) $E_{nm}= 1.598 \MeV (1/m^2
-1/n^2)$ (or from the similar transitions corresponding to the case
$I_o = 1.287 \MeV $) at the level $3 \cdot 10^{-4}S_3^{-2}(\cm^2 \s
\MeV ster)^{-1}$.
\subsection{O-helium solution for dark matter puzzles}
It should be noted that the nuclear cross section of the O-helium
interaction with matter escapes the severe constraints\cite{McGuire:2001qj,McGuire2,ZF}
on strongly interacting dark matter particles
(SIMPs)\cite{Starkman,Wolfram,Starkman2,Javorsek,Mitra,Mack,McGuire:2001qj,McGuire2,ZF} imposed by the XQC experiment\cite{XQC,XQC1}. Therefore, a special strategy of direct O-helium  search
is needed, as it was proposed in\cite{Belotsky:2006fa}.

\subsubsection{O-helium in the terrestrial matter} The evident
consequence of the O-helium dark matter is its inevitable presence
in the terrestrial matter, which appears opaque to O-helium and
stores all its in-falling flux.

After they fall down terrestrial surface, the in-falling $OHe$
particles are effectively slowed down due to elastic collisions with
matter. Then they drift, sinking down towards the center of the
Earth with velocity \beq V = \frac{g}{n \sigma v} \approx 80 S_3
A_{med}^{1/2} \cm/\s. \label{dif}\eeq Here $A_{med} \sim 30$ is the average
atomic weight in terrestrial surface matter, $n=2.4 \cdot 10^{24}/A$
is the number of terrestrial atomic nuclei, $\sigma v$ is the rate
of nuclear collisions and $g=980~ \cm/\s^2$.

Near the Earth's surface, the O-helium abundance is determined by
the equilibrium between the in-falling and down-drifting fluxes.

At a depth $L$ below the Earth's surface, the drift timescale is
$t_{dr} \sim L/V$, where $V \sim 400 S_3 \cm/\s$ is the drift velocity and $m_o=S_3 \TeV$ is the mass of O-helium. It means that the change of the incoming flux,
caused by the motion of the Earth along its orbit, should lead at
the depth $L \sim 10^5 \cm$ to the corresponding change in the
equilibrium underground concentration of $OHe$ on the timescale
$t_{dr} \approx 2.5 \cdot 10^2 S_3^{-1}\s$.

The equilibrium concentration, which is established in the matter of
underground detectors at this timescale, is given by
\begin{equation}
    n_{oE}=n_{oE}^{(1)}+n_{oE}^{(2)}\cdot sin(\omega (t-t_0))
    \label{noE}
\end{equation}
with $\omega = 2\pi/T$, $T=1yr$ and
$t_0$ the phase.
So, there is a averaged concentration given by
\begin{equation}
    n_{oE}^{(1)}=\frac{n_o}{320S_3 A_{med}^{1/2}} V_{h}
\end{equation}
and the annual modulation of concentration characterized by the amplitude
\begin{equation}
    n_{oE}^{(2)}= \frac{n_o}{640S_3 A_{med}^{1/2}} V_E.
\end{equation}
Here $V_{h}$-speed of Solar System (220 km/s), $V_{E}$-speed of
Earth (29.5 km/s) and $n_{0}=3 \cdot 10^{-4} S_3^{-1} \cm^{-3}$ is the
local density of O-helium dark matter.

\subsubsection{OHe in the underground detectors}

The explanation\cite{Levels,DMDA,DADM} of the results of
DAMA/NaI\cite{DAMA-review} and DAMA/LIBRA\cite{Bernabei:2008yi}
(see Ref. \citen{DAMARev} for the latest review of these results)
experiments is based on the idea that OHe,
slowed down in the matter of detector, can form a few keV bound
state with nucleus, in which OHe is situated \textbf{beyond} the
nucleus. Therefore the positive result of these experiments is
explained by annual modulation in reaction of radiative capture of OHe
\begin{equation}
A+(^4He^{++}O^{--}) \rightarrow [A(^4He^{++}O^{--})]+\gamma
\label{HeEAZ}
\end{equation}
by nuclei in DAMA detector.

To simplify the solution of Schrodinger equation the
potential was approximated in Refs. \citen{Levels,Levels1} by a rectangular potential, presented on Fig.~\ref{pic1}.
Solution of Schrodinger equation determines the condition, under
which a low-energy  OHe-nucleus bound state appears in the shallow well of the region
III and the range of nuclear parameters was found, at which OHe-sodium binding energy is in the interval 2-4 keV.


The rate of radiative capture of OHe by nuclei can be calculated\cite{Levels,DMDA}
with the use of the analogy with the radiative
capture of neutron by proton with the account for: i) absence of M1
transition that follows from conservation of orbital momentum and
ii) suppression of E1 transition in the case of OHe. Since OHe is
isoscalar, isovector E1 transition can take place in OHe-nucleus
system only due to effect of isospin nonconservation, which can be
measured by the factor $f = (m_n-m_p)/m_N \approx 1.4 \cdot
10^{-3}$, corresponding to the difference of mass of neutron,$m_n$,
and proton,$m_p$, relative to the mass of nucleon, $m_N$. In the
result the rate of OHe radiative capture by nucleus with atomic
number $A$ and charge $Z$ to the energy level $E$ in the medium with
temperature $T$ is given by
\begin{equation}
    \sigma v=\frac{f \pi \alpha}{m_p^2} \frac{3}{\sqrt{2}} (\frac{Z}{A})^2 \frac{T}{\sqrt{Am_pE}}.
    \label{radcap}
\end{equation}

Formation of OHe-nucleus bound system leads to energy release of its
binding energy, detected as ionization signal.  In the context of
our approach the existence of annual modulations of this signal in
the range 2-6 keV and absence of such effect at energies above 6 keV
means that binding energy $E_{Na}$ of Na-OHe system in DAMA experiment should
not exceed 6 keV, being in the range 2-4 keV. The amplitude of
annual modulation of ionization signal can reproduce the result of DAMA/NaI and DAMA/LIBRA
experiments for $E_{Na} = 3 \keV$. The
account for energy resolution in DAMA experiments\cite{DAMAlibra}
can explain the observed energy distribution of the signal from
monochromatic photon (with $E_{Na} = 3 \keV$) emitted in OHe
radiative capture.

At the corresponding nuclear parameters there is no binding
of OHe with iodine and thallium\cite{Levels}.

It should be noted that the results of DAMA experiment exhibit also
absence of annual modulations at the energy of MeV-tens MeV. Energy
release in this range should take place, if OHe-nucleus system comes
to the deep level inside the nucleus. This transition implies
tunneling through dipole Coulomb barrier and is suppressed below the
experimental limits.

For the chosen range of nuclear parameters, reproducing the results
of DAMA/NaI and DAMA/LIBRA, the results of Ref. \citen{Levels} indicate that
there are no levels in the OHe-nucleus systems for heavy nuclei. In
particular, there are no such levels in Xe, what
seem to prevent direct comparison with DAMA results in
XENON100 experiment\cite{xenon}. The existence of such level in Ge and the comparison with the results of
CDMS\cite{CDMS,CDMS2,CDMS3} and CoGeNT\cite{cogent} experiments need special study. According to Ref. \citen{Levels} OHe should bind with O and Ca, what is of interest for interpretation of the signal, observed in CRESST-II experiment\cite{cresst}.

In the thermal equilibrium OHe capture rate is proportional to the temperature. Therefore it looks
like it is suppressed in cryogenic detectors by a factor of order
$10^{-4}$. However, for the size of cryogenic devices  less, than
few tens meters, OHe gas in them has the thermal velocity of the
surrounding matter and this velocity dominates in the relative velocity of OHe-nucleus system.
It gives the suppression relative to room temperature
only $\sim m_A/m_o$. Then the rate of OHe radiative capture in
cryogenic detectors is given by Eq.(\ref{radcap}), in which room
temperature $T$ is multiplied by factor $m_A/m_o$. Note that in the case of $T=70\K$ in CoGeNT experiment
relative velocity is determined by the thermal velocity of germanium nuclei, what leads to enhancement relative to cryogenic germanium detectors.
\subsection{Conclusions}
The existence of heavy stable particles is one of the popular solutions for the dark matter problem.
Usually they are considered to be electrically neutral. But potentially dark matter can be formed by
stable heavy charged particles bound in neutral atom-like states by Coulomb attraction.
Analysis of the cosmological data and atomic composition of the Universe gives the constrains
on the particle charge showing that  only $-2$
charged constituents, being trapped by primordial helium
in neutral O-helium states, can avoid the problem of overproduction of the anomalous isotopes of chemical elements, which are severely constrained by observations. Cosmological model of O-helium dark matter
can even explain puzzles of direct dark matter searches.

The proposed explanation is based on the mechanism of low energy
binding of OHe with nuclei. Within the uncertainty of nuclear
physics parameters there exists a range at which OHe binding energy
with sodium is in the interval 2-4 keV. Annual modulation in radiative capture of OHe to
this bound state leads to the corresponding energy release observed
as an ionization signal in DAMA/NaI and
DAMA/LIBRA experiments.


With the account for high sensitivity of the numerical results to
the values of nuclear parameters and for the approximations, made in
the calculations, the presented results can be considered only as an
illustration of the possibility to explain puzzles of dark matter
search in the framework of composite dark matter scenario. An
interesting feature of this explanation is a conclusion that the
ionization signal may
be absent in detectors containing light (e.g. $^3He$) or heavy (e.g. Xe) elements.
Therefore test of results of DAMA/NaI and
DAMA/LIBRA experiments by other experimental groups can become a
very nontrivial task. Recent indications to positive result in the matter of CRESST detector\cite{cresst},
in which OHe binding is expected together with absence of signal in xenon detector\cite{xenon}, may qualitatively favor the presented approach. For the same chemical content
an order of magnitude suppression in cryogenic detectors can explain why indications to positive effect in
CoGeNT experiment\cite{cogent} can be compatible with the constraints of CDMS experiment.

The present explanation contains distinct
features, by which it can be distinguished from
other recent approaches to this problem\cite{Edward,Foot,Feng1,Feng2,Drob,Feldstein:2009tr,Bai,Feldstein:2009np,Fitzpatrick:2010em,Andreas:2010dz,Alves:2010dd,Barger:2010yn,Savage:2010tg,Hooper:2010uy,Chang:2010pr,Chang:2010en,Barger:2010gv,Feldstein:2010su}

An inevitable consequence of the proposed explanation is appearance
in the matter of underground detectors anomalous
superheavy isotopes, having the mass roughly by $m_o$
larger, than ordinary isotopes of the corresponding elements.

It is interesting to note that in the framework of the presented approach
positive result of experimental search for WIMPs by effect of their
nuclear recoil would be a signature for a multicomponent nature of
dark matter. Such OHe+WIMPs multicomponent dark matter scenarios
naturally follow from AC model \cite{FKS} and can be realized in
models of Walking technicolor \cite{KK2}.

Stable $-2$ charge states ($O^{--}$) can be elementary like AC-leptons or technileptons,
or look like technibaryons. The latter, composed of techniquarks, reveal their structure at much higher energy scale and should be produced at LHC as
elementary species. The signature  for AC leptons and techniparticles is unique and distinctive what  allows
to separate them  from other hypothetical exotic particles.

Since simultaneous production of three $U \bar U$ pairs and
their conversion in two doubly charged quark clusters $UUU$
is suppressed, the only possibility to test the
models of composite dark matter from 4th generation in the collider experiments is a search for production of stable hadrons containing single $U$ or $\bar U$ like $Uud$ and $\bar U u$/$\bar U d$.

The presented approach sheds new light on the physical nature of
dark matter. Specific properties of dark atoms and their
constituents are challenging for the experimental search. The
development of quantitative description of OHe interaction with
matter confronted with the experimental data will provide the
complete test of the composite dark matter model. It challenges search for stable double charged particles at accelerators and cosmic rays as direct experimental probe for charged constituents of dark atoms of dark matter.
\section{Discussion}\label{Discussion}

Observational cosmology offers strong evidences favoring the
existence of processes, determined by new physics, and the
experimental physics approaches to their investigation.
Cosmoparticle physics \cite{ADS,MKH,book,book3}, studying the
physical, astrophysical and cosmological impact of new laws of
Nature, explores the new forms of matter and their physical
properties. Its development offers the great challenge for
theoretical and experimental research. Physics of dark matter in all its aspects
plays important role in this process.

The new physics follows from the necessity to extend the Standard
model. The white spots in the representations of symmetry groups,
considered in the extensions of the Standard model, correspond to
new unknown particles. The extension of the symmetry of gauge
group puts into consideration new gauge fields, mediating new
interactions. Global symmetry breaking results in the existence of
Goldstone boson fields.

For a long time the necessity to extend the Standard model had
purely theoretical reasons. Aesthetically, because full
unification is not achieved in the Standard model; practically,
because it contains some internal inconsistencies. It does not
seem complete for cosmology. One has to go beyond the Standard
model to explain inflation, baryosynthesis and nonbaryonic dark
matter. The discovery of neutrino oscillations (see for
review e.g. Ref. \citen{Jung}) and the experimental
evidences for the existence of
dark matter particles \cite{DAMA-review} indicate that the experimental searches may have
already crossed the border of new physics.

In particle physics direct experimental probes for the predictions
of particle theory are most attractive. The predictions of new
charged particles, such as supersymmetric particles or quarks and
leptons of new generation, are accessible to experimental search
at accelerators of new generation, if their masses are in
100GeV-1TeV range. However, the predictions related to higher
energy scale need non-accelerator or indirect means for their
test.

The search for rare processes, such as proton decay, neutrino
oscillations, neutrinoless beta decay, precise measurements of
parameters of known particles, experimental searches for dark
matter represent the widely known forms of such means.
Cosmoparticle physics offers the nontrivial extensions of indirect
and non-accelerator searches for new physics and its possible
properties. In experimental cosmoarcheology the data is to be
obtained, necessary to link the cosmophenomenology of new physics
with astrophysical observations (See Ref. \citen{Cosmoarcheology}). In
experimental cosmoparticle physics the parameters, fixed from the
consitency of cosmological models and observations, define the
level, at which the new types of particle processes should be
searched for (see Ref. \citen{expcpp}).

The theories of everything should provide the complete physical
basis for cosmology. The problem is that the string theory
\cite{Green} is now in the form of "theoretical theory", for which
the experimental probes are widely doubted to exist. The
development of cosmoparticle physics can remove these doubts. In
its framework there are two directions to approach the test of
theories of everything.

One of them is related with the search for the experimentally
accessible effects of heterotic string phenomenology. The
mechanism of compactification and symmetry breaking leads to the
prediction of homotopically stable objects \cite{Kogan1} and
shadow matter \cite{Kogan2}, accessible to cosmoarcheological
means of cosmoparticle physics. The condition to reproduce the
Standard model naturally leads in the heterotic string
phenomenology to the prediction of fourth generation of quarks and
leptons \cite{Shibaev} with a stable massive 4th neutrino
\cite{Fargion99}, what can be the subject of complete experimental
test in the near future. The comparison between the rank of the
unifying group $E_{6}$ ($r=6$) and the rank of the Standard model
($r=4$) implies the existence of new conserved charges and new
(possibly strict) gauge symmetries. New strict gauge U(1) symmetry
(similar to U(1) symmetry of electrodynamics) is possible, if it
is ascribed to the fermions of 4th generation. This hypothesis
explains the difference between the three known types of neutrinos
and neutrino of 4th generation. The latter possesses new gauge
charge and, being Dirac particle, can not have small Majorana mass
due to sea saw mechanism. If the 4th neutrino is the lightest
particle of the 4th quark-lepton family, strict conservation of
the new charge makes massive 4th neutrino to be absolutely stable.
Following this hypothesis \cite{Shibaev} quarks and leptons of 4th
generation are the source of new long range interaction
($y$-electromagnetism), similar to the electromagnetic interaction
of ordinary charged particles. If proved, the
practical importance of this property could be hardly
overestimated.

It is interesting, that heterotic string phenomenology  embeds
even in its simplest implementation both supersymmetric particles and
the 4th family of quarks and leptons, in particular, the two types
of WIMP candidates: neutralinos and massive stable 4th neutrinos, as well as nuclear interacting OHe dark atoms,
built up by stable (anti-)U quarks of 4th generation.
So in the framework of this phenomenology the multicomponent
analysis of WIMP effects is favorable.

In the above approach some particular phenomenological features of
simplest variants of string theory are studied. The other
direction is to elaborate the extensive phenomenology of theories
of everything by adding to the symmetry of the Standard model the
(broken) symmetries, which have serious reasons to exist. The
existence of (broken) symmetry between quark-lepton families, the
necessity in the solution of strong CP-violation problem with the
use of broken Peccei-Quinn symmetry, as well as the practical
necessity in supersymmetry to eliminate the quadratic divergence
of Higgs boson mass in electroweak theory is the example of
appealing additions to the symmetry of the Standard model. The
horizontal unification and its cosmology represent the first step
on this way, illustrating the approach of cosmoparticle physics to
the elaboration of the proper phenomenology for theories of
everything \cite{Sakharov1}.

For long time scenarios with Primordial Black holes belonged
dominantly to cosmological {\it anti-Utopias}, to "fantasies", which
provided restrictions on physics of very early Universe from
contradiction of their predictions with observational data. Even
this "negative" type of information makes PBHs an important
theoretical tool. Being formed in the very early Universe as
initially nonrelativistic form of matter, PBHs should have increased
their contribution to the total density during RD stage of
expansion, while effect of PBH evaporation should have strongly
increased the sensitivity of astrophysical data to their presence.
It links astrophysical constraints on hypothetical sources of cosmic
rays or gamma background, on hypothetical factors, causing influence
on light element abundance and spectrum of CMB, to restrictions on
superheavy particles in early Universe and on first and second order
phase transitions, thus making a sensitive astrophysical probe to
particle symmetry structure and pattern of its breaking at superhigh
energy scales.

Gravitational mechanism of particle creation in PBH evaporation
makes evaporating PBH an unique source of any species of particles,
which can exist in our space-time. At least theoretically, PBHs can
be treated as source of such particles, which are strongly
suppressed in any other astrophysical mechanism of particle
production, either due to a very large mass of these species, or
owing to their superweak interaction with ordinary matter.

By construction astrophysical constraint excludes effect, predicted
to be larger, than observed. At the edge such constraint converts
into an alternative mechanism for the observed phenomenon. At some
fixed values of parameters, PBH spectrum can play a positive role
and shed new light on the old astrophysical problems.

The common sense is to think that PBHs should have small sub-stellar
mass. Formation of PBHs within cosmological horizon, which was very
small in very early Universe, seem to argue for this viewpoint.
However, phase transitions on inflationary stage can provide spikes
in spectrum of fluctuations at any scale, or provide formation of
closed massive domain walls of any size.

In the latter case primordial clouds of massive black holes around
intermediate mass or supermassive black hole is possible. Such
clouds have a fractal spatial distribution. A development of this
approach gives ground for a principally new scenario of the galaxy
formation in the model of the Big Bang Universe. Traditionally, Big
Bang model assumes a homogeneous distribution of matter on all
scales, whereas the appearance of observed inhomogeneities is
related to the growth of small initial density perturbations.
However, the analysis of the cosmological consequences of the
particle theory indicates the possible existence of strongly
inhomogeneous primordial structures in the distribution of both the
dark matter and baryons. These primordial structures represent a new
factor in galaxy formation theory. Topological defects such as the
cosmological walls and filaments, primordial black holes, archioles
in the models of axionic CDM, and essentially inhomogeneous
baryosynthesis (leading to the formation of antimatter domains in
the baryon-asymmetric Universe
\cite{exl1,exl2,crg,kolb,we,khl,CSKZ,zil,sb,dolgmain,
Golubkov,ANTIHE,ANTIME,book,book2,book3}) offer by no means a
complete list of possible primary inhomogeneities inferred from the
existing elementary particle models.

We can conclude that from the very beginning to the modern stage,
the evolution of Universe is governed by the forms of matter,
different from those we are built of and observe around us. From
the very beginning to the present time, the evolution of the
Universe was governed by physical laws, which we still don't know.
These laws follow from the fundamental particle symmetry beyond the Standard model.
Observational cosmology offers strong evidences favoring the
existence of processes, determined by such laws of new physics, and the
experimental physics approaches to their investigation.

Cosmoparticle physics originates from the well established
relationship between microscopic and macroscopic descriptions in
theoretical physics. Remind the links between statistical physics
and thermodynamics, or between electrodynamics and theory of
electron. To the end of the XX Century the new level of this
relationship was realized. It followed both from the cosmological
necessity to go beyond the world of known elementary particles in
the physical grounds for inflationary cosmology with
baryosynthesis and dark matter as well as from the necessity for
particle theory to use cosmological tests as the important and in
many cases unique way to probe its predictions.

Cosmoparticle physics \cite{ADS} \cite{MKH}, studying the
physical, astrophysical and cosmological impact of new laws of
Nature, explores the new forms of matter and their physical
properties, what opens the way to use the corresponding new
sources of energy and new means of energy transfer. It offers the
great challenge for the new Millennium.


\end{document}